\documentclass[useAMS,usenatbib]{mn2e}
\usepackage{epsfig}

\newcommand{\prob}{{\mathcal P}}
\newcommand{\av}[1]{\langle #1\rangle}
\newcommand{\avv}[1]{\langle\langle #1\rangle\rangle}
\newcommand{\avb}[1]{\overline{#1}}

\title[Analysis of Radial Velocities]{An Integrated Analysis of Radial Velocities in Planet Searches}
\author[A. Cumming and D. Dragomir]{Andrew Cumming$^{1}$ and Diana Dragomir$^{1,2}$\\
$^{1}$Department of Physics, McGill University, 3600 rue University, Montreal, QC H3A 2T8, Canada\\ $^{2}$Department of Physics and Astronomy, University of British Columbia, 
6224 Agricultural Road, Vancouver, BC V6T1Z1, Canada}

\begin{document}

\date{}

\pagerange{\pageref{firstpage}--\pageref{lastpage}} \pubyear{2009}

\maketitle

\label{firstpage}

\begin{abstract}
We discuss a Bayesian approach to the analysis of radial velocities in planet searches. We use a combination of exact and approximate analytic and numerical techniques to efficiently evaluate $\chi^2$ for multiple values of orbital parameters, and to carry out the marginalization integrals for a single planet including the possibility of a long term trend. The result is a robust algorithm that is rapid enough for use in real time analysis that outputs constraints on orbital parameters and false alarm probabilities for the planet and long term trend. The constraints on parameters and odds ratio that we derive compare well with previous calculations based on Markov Chain Monte Carlo methods, and we compare our results with other techniques for estimating false alarm probabilities and errors in derived orbital parameters. False alarm probabilities from the Bayesian analysis are systematically higher than frequentist false alarm probabilities, due to the different accounting of the number of trials. We show that upper limits on the velocity amplitude derived for circular orbits are a good estimate of the upper limit on the amplitude of eccentric orbits for $e\la 0.5$.
\end{abstract}



\section{Introduction}

The analysis of a set of radial velocities in planet searches typically involves a number of different steps. First, the best fitting Keplerian orbital parameters are found by minimizing $\chi^2$, for example with a Levenberg-Marquardt algorithm \citep{nr}. Because of the complex multimodal shape of the $\chi^2$ distribution in parameter space, a Lomb-Scargle periodogram \citep{lomb,scargle} is often used beforehand to fit circular orbits at a range of orbital periods, providing starting points for the Keplerian fit. Then the reality of the signal is assessed by calculating the false alarm probability (FAP) that the observed signal could arise due to noise fluctuations, typically using Monte Carlo simulations  \citep{marcy05,c04}. This has become particularly important as radial velocity surveys reveal planets with lower velocity amplitudes, comparable to the measurement uncertainties and other sources of noise. A related question is comparing different models for the data, for example deciding whether a two (or more) planet model is preferred over a single planet model, or whether to include a long term trend due to a long period companion (see for example, \citealt{robinson}). 

Uncertainties in the fitted orbital parameters are then calculated. A common technique is to scramble the residuals to the best fit Keplerian orbit, add them back to the predicted velocity curve, and refit the orbit. After repeating this many times, the distribution of fitted parameters gives an estimate of the uncertainty \citep{marcy05}. Another approach is to use Bayesian methods implemented with Markov Chain Monte Carlo (MCMC) simulations \citep{ford05}. In the case of a non-detection, the upper limit on the planet mass as a function of orbital period is an important input for population studies \citep{walker,c99,endl02,wittenmyer06}. 

Often, all of these steps must be carried out for a given radial velocity data set. Many of them are based on Monte Carlo simulations involving fitting Keplerian orbits to synthetic data sets. These trials can become cumbersome for the large numbers of orbital frequencies that must be considered. For this reason, recent calculations have produced upper limits for circular orbits only \citep{c08}, relying on the fact that detectability of planets falls off with eccentricity only for $e\ga 0.6$ \citep{endl02,c04}, or on a sparse grid of orbital period values (O' Toole et al. 2008).

We focus in this paper on a Bayesian approach to the analysis of radial velocity data. The advantage is that, in principle, a Bayesian analysis answers all of the above questions with a single calculation, providing constraints on model parameters and odds ratios which can be used to decide which model best describes the data \citep{ford05,gregory05,c04}. This would simplify analysis of radial velocity data sets. The difficulty in practice is that the marginalization over parameters requires the evaluation of multidimensional integrals over parameter space.

Bayesian methods have been applied to planet searches, using sophisticated Markov Chain Monte Carlo (MCMC) techniques to evaluate the integrals. \cite{ford05} applied this technique to determining the constraints on orbital parameters, while Gregory in a series of papers \citep{gregory05,gregory07a,gregory07b} also considers model comparison. \cite{ford06} investigated different proposal distribution functions to help speed convergence of MCMC chains. Whereas the chains used by \cite{ford05} were $10^6$--$10^{10}$ steps in length, \cite{ford06} found that it was possible to achieve convergence after only $10^4$--$10^6$ steps by optimizing the directions in parameter space in which steps are taken. Recently, \cite{balan} have developed an MCMC code {\em Exofit} based on the methodology of \cite{fordgregory07} which is publicly available\footnote{Available at  http://www.http://zuserver2.star.ucl.ac.uk/~lahav/exofit.html .}. 

Despite this tremendous progress, the application of MCMC to radial velocity data is not yet routine, although it is commonly used to assess uncertainties in orbital parameters. As well as optimizing the steps in parameter space, one important difficulty in using the MCMC approach is assessing whether the chains have converged. Another is that when the signal to noise ratio is low and the distribution of $\chi^2$ in parameter space is multimodal, the MCMC chain may miss minima in $\chi^2$. \cite{gregory05} introduced a parallel tempering scheme in which several MCMC chains are run simultaneously, each with a different temperature, hotter chains making larger jumps in parameter space, colder chains exploring local minima. As the calculation progresses, the chains exchange information in a way  that preserves their statistical character. This scheme has been successfully applied to multiple planet systems \citep{gregory07a,gregory07b}.

In this paper, we take a different approach. We consider models with one planet only, or one planet plus a long term linear trend, and use a combination of grid-based numerical evaluation and exact and approximate analytic methods to evaluate the marginalization integrals. The idea is to look for ways in which the marginalization integrals can be evaluated more efficiently. As well as providing a useful tool for analysing radial velocity data for single planet systems,  it provides a check on the output of MCMC simulations, and may have application to making MCMC codes for analysis of multiple planet systems more efficient.

We start in \S 2 with an overview of the Bayesian approach, including how to write down the posterior probabilities for orbital parameters, and how to use them to calculate false alarm probabilities. In \S 3, we discuss circular orbits, using analytic techniques to evaluate the marginalization integrals. In \S 4, we divide the parameters for eccentric orbits into fast (linear) and slow (non-linear) parameters, and use the analytic techniques for circular orbits to marginalize over the fast parameters. In \S 5, we compare our results to MCMC calculations, and traditional methods for evaluating false alarm probabilities and upper limits on companion mass.


\section{Overview}
\label{sec:overview}

Bayesian analysis of radial velocity data has been discussed previously by several authors \citep{ford05,ford06,ford08,fordgregory07,gregory05a,gregory05,gregory07a,gregory07b,c04}. Here we give a brief reminder of the basic ideas and introduce our notation, and show how the systemic velocity and noise uncertainty can be analytically marginalized.

\subsection{Parameter estimation}

We start with a model for the radial velocities with set of parameters $\vec{a}$. For example, a single Keplerian orbit has six parameters $\vec{a}=(K,P,e,\omega_p,t_p,\gamma)$, where $K$ is the velocity amplitude, $P$ the orbital period, $e$ the eccentricity, $\omega_p$ and $t_p$ are the argument and time of pericenter, and $\gamma$ is the systemic velocity. The data consist of a set of $N$ measured velocities $v_i$, observation times $t_i$, and errors $\sigma_i$. Bayes' theorem allows us to calculate the probability distribution of the parameters $\vec{a}$ given the data, also known as the posterior probability of $\vec{a}$,
\begin{equation}\label{eq:bayes}
\prob\left(\vec{a}|d\right)={\prob(\vec{a})\prob(d|\vec{a})\over \prob(d)},
\end{equation}
where $\prob(d)$ is a normalization factor. The term $\prob(\vec{a})$ is the prior probability distribution for the parameters $\vec{a}$, which allows us to specify any knowledge of the parameter distribution that we have before the data are taken.
If the errors are Gaussian-distributed and uncorrelated, the likelihood of the data, or probability of the data given a particular choice of model parameters is
\begin{equation}\label{eq:likelihood}
\prob\left(d|\vec{a}\right)={1\over \prod_i (2\pi)^{1/2}\sigma_i}\exp\left({-{\chi^2(\vec{a})\over 2}}\right)
\end{equation}
where
\begin{equation}
\chi^2(\vec{a})=\sum_{i=1}^Nw_i\left(v_i-V_i(\vec{a})\right)^2
\end{equation}
is the usual $\chi^2$ statistic, written in terms of weights $w_i=1/\sigma_i^2$.  We write the model velocity at time $t_i$ as $V_i$.

Often, we are interested in the probability distribution of a single parameter, or a subset of parameters. For example, a circular orbit has $\vec{a}=(K,P,\gamma,\phi)$, where $K$ is the velocity amplitude, $P$ the orbital period, $\gamma$ the systemic velocity, and $\phi$ the orbital phase. It is likely that we are not interested in the particular values of $\gamma$ or $\phi$, but want to constrain the orbital period and velocity amplitude. The joint probability distribution for $P$ and $K$ can be obtained by marginalizing over the other parameters, 
\begin{equation}
\prob(P,K|d)=\int d\phi \int d\gamma\ \prob(P,K,\phi,\gamma|d).
\end{equation}
Marginalization amounts to performing a weighted average of the probability distribution over the unwanted parameters. 

A number of other useful quantities can be obtained from $\prob(P,K|d)$. Further integration over $K$ gives $\prob(P|d)$, or integration over $P$ gives $\prob(K|d)$. A confidence interval for $K$ can be calculated from $\prob(K|d)$. For example, if a planet is not detected in a given data set, an upper limit can be placed on the amplitude of undetected orbits. The 99\% upper limit $K_{99}$ is given by $\int_0^{K_{99}} dK \prob(K|d)/\int_0^\infty dK \prob(K|d) = 0.99$.

For eccentric orbits, we will focus in this paper on obtaining $\prob(P,e,K|d)$ by marginalizing over $\gamma$, $\omega_p$, and $t_p$.

\subsection{The noise distribution}

In equation (\ref{eq:likelihood}), we assumed that the standard deviation of the noise for each observation $\sigma_i$ was given. In reality, other noise sources may be present in the data that hinder the identification of planetary signals, for example intrinsic stellar ``jitter'' (e.g.~\citealt{wright05}) due to rotation of spots across the surface of the star, or changes in line profiles over time related to magnetic activity. This extra noise can be incorporated as an additional parameter of the model. A common choice (\citealt{gregory05,ford06}) is to add the extra noise term in quadrature with the measurement errors $\sigma_i$. 

Here, we instead multiply each value of $\sigma_i$ by a noise scaling factor $k$ \citep{c04,gregory05a}, and analytically marginalize over $k$ (e.g.~\citealt{sivia})
\begin{equation}\label{eq:noisemarg}
\prob(d|\vec{a})\propto \int_0^\infty {dk\over k} {1\over k^N}\exp\left(-{\chi^2(\vec{a})\over 2k^2}\right)\propto \left(\chi^2(\vec{a})\right)^{-N/2}.
\end{equation}
The constant prefactor, which depends only on the weights $w_i$ and the number of observations $N$, does not affect the shape of the posterior probability distributions, and cancels out when we calculate odds ratios. Therefore, we drop it and replace equation (\ref{eq:likelihood}) with
\begin{equation}\label{eq:likelihood2}
\prob(d|\vec{a})=\left(\chi^2(\vec{a})\right)^{-N/2}.
\end{equation}
This is a Student's t-distribution rather than Gaussian distribution \citep{sivia}.

We take an infinite range for $k$ in equation (\ref{eq:noisemarg}), whereas in reality we likely have some information about the uncertainty in the noise level. For example, we may be able to estimate the size of the expected stellar jitter based on stellar properties \citep{wright05}. Alternatively, we could keep $k$ as a parameter, and evaluate the constraints on $k$ from the data, $\prob(k|d)$ (e.g. \citealt{ford06}). We have tried marginalizing numerically over $k$ with finite limits, and find that the results for realistic ranges of $k$ are close to the analytic case with infinite limits. Therefore we marginalize analytically over $k$ and adopt equation (\ref{eq:likelihood2}) as the likelihood throughout this paper\footnote{The techniques we develop below for rapidly marginalizing over parameters can also be applied to the case where $k$ is kept as a parameter. This is discussed in Appendix B.}.

\subsection{Priors}

The choice of appropriate prior probabilities $\prob(\vec{a})$ for the various parameters has been discussed in depth in the literature (e.g.~\citealt{gregory05,fordgregory07}). We mostly follow this previous work. For circular orbits, we use uniform priors for $\gamma$ and $\phi$, and priors for $K$,$P$ that are uniform in log (the Jeffreys prior). For eccentric orbits, we take uniform priors in $\gamma$, $t_p$, $\omega_p$, $e$, and log-uniform priors in $K$, and $P$, i.e.~$\prob(P)=1/(P\log(P_2/P_1))$ and $\prob(K)=1/(K\log(K_2/K_1))$. If a long term trend is included in the model (a linear term $\beta t_i$; see Appendix A), we take a uniform prior in the slope $\beta$. 

In fact, \cite{gregory05} and \cite{fordgregory07} use a modified Jeffreys prior for the noise term and the velocity amplitude rather than the standard Jeffreys prior. A modified Jeffreys prior is uniform in log above some scale, and uniform below that scale. For radial velocity amplitude $K$ or extra noise term, the turnover scale is taken to be $\approx 1\ {\rm m/s}$. The values of $K$ we are interested in are typically larger than this, and so for simplicity we use a Jeffreys prior between our lower and upper limits in $K$. The ranges that we take are $K=1\ {\rm m/s}$ to $2\Delta v$, where $\Delta v$ is the observed range of velocities, and $P=1$ day to the time span of the observations.

\subsection{Model comparison and the false alarm probability}

Marginalizing over all the parameters of a model gives the total probability of that model. For example, given $\prob(P,K|d)$ for circular orbits, we could calculate the total probability that a planet is present
\begin{equation}
\prob(1|d)=\int dP \int dK\ \prob(P,K|d).
\end{equation}
Similarly, by considering a model without a planet, we can calculate the probability that no planet is present given the data, $\prob(0|d)$. We define the normalization $\prob(d)$ in equation (\ref{eq:bayes}) so that the sum over the probabilities of all models is unity. For example, if we consider only two possible models, that there is or is not a planet present, we choose $\prob(d)$ such that 
\begin{equation}
\prob(1|d)+\prob(0|d)=1.
\end{equation}

We can think of the posterior probability that there is no planet present $\prob(0|d)$ as the false alarm probability. It can be written without including the $\prob(d)$ factors explicitly as
\begin{equation}\label{eq:F}
F=\prob(0|d)={1\over 1+\Lambda}
\end{equation}
where $\Lambda$ is the odds ratio
\begin{equation}
\Lambda={\prob(1|d)\over \prob(0|d)}
\end{equation}
(the normalization factors $\prob(d)$ cancel out when the ratio is taken).
For $\Lambda\gg 1$, $F\approx\Lambda^{-1}$. The odds ratio is
\begin{equation}\label{eq:oddscirc}
\Lambda={\int dK \int dP\ \prob(P,K|d)\over \prob(0|d)},
\end{equation}
for circular orbits,
or
\begin{equation}\label{eq:oddsecc}
\Lambda={\int dK \int dP \int de\ \prob(P,K,e|d)\over \prob(0|d)},
\end{equation}
for eccentric orbits. 

This approach can be generalized to more than two models. For example, later we will consider four possible models for a given star, the possible combinations of including or not including a Keplerian orbit with period less than the time span of the observations, and including or not including a long term trend. To calculate the false alarm probability associated with the short period planet, we define the odds ratio
\begin{equation}\label{eq:oddswithtrend}
\Lambda={\prob(1|d)+\prob(1,t|d)\over \prob(0|d)+\prob(0,t|d)}
\end{equation}
where $1$ or $0$ indicate that the short period planet is or is not included in the model, and $t$ indicates that a long term trend is included.

\subsection{Probability that there is no planet $\prob(0|d)$}

The posterior probability of no planet $\prob(0|d)$, where the ``no planet'' model is a constant velocity $V_i=\gamma$, can be calculated analytically. Using the likelihood of equation (\ref{eq:likelihood2})
\begin{equation}
\prob(0|d)={1\over \prob(d)\Delta\gamma}\int_{\gamma_1}^{\gamma_2}\ d\gamma\ \left(\chi^2(\gamma)\right)^{-N/2}
\end{equation}
where $\Delta\gamma=\gamma_2-\gamma_1$ is the range of values of $\gamma$ considered, and we assume a uniform prior for $\gamma$ in that range. Minimizing $\chi^2$ with respect to $\gamma$, we find the best-fitting value $\gamma_0=\sum w_i v_i/\sum w_i$. In terms of $\gamma_0$, we can write
\begin{equation}\label{eq:gammachi}
\chi^2\left(\gamma\right)=\chi^2(\gamma_0)+(\gamma-\gamma_0)^2\sum w_i.
\end{equation}
The fact that the distribution of $\chi^2(\gamma)$ is analytic is mentioned in \cite{ford06}. For clarity, we drop the subscript $i$ on the sum in equation (\ref{eq:gammachi}) and in the remainder of the paper, a sum over the observations with $i$ running from $1$ to $N$ is implied.

The quadratic form of $\chi^2$ allows the integral over $\gamma$ to be carried out analytically when the limits $\gamma_1\rightarrow -\infty$ and $\gamma_2\rightarrow\infty$. In that limit, the normalization factor diverges, $\Delta\gamma\rightarrow\infty$. However the values of $\Delta\gamma$ cancel when we form an odds ratio, as does the normalization factor $\prob(d)$. Therefore, we can drop the prefactor after integrating, giving the final result
\begin{equation}
\prob(0|d)=\left(\chi^2(\gamma_0)\right)^{-(N-1)/2}.
\end{equation}

An alternative ``no planet'' model is a linear trend in the radial velocities over time, $V_i=\gamma+\beta t_i$. A linear term is often included (and needed) in radial velocity fits to account for additional companions with long orbital periods. A similar formula for $\prob(0|d)$ can be derived in that case. For clarity, we leave this to Appendix A, along with how to add a linear term to the circular and Keplerian orbit fits, and consider only the constant velocity no planet model in the main text.


\section{Circular orbits}

In the previous section, we saw that a calculation of $\prob(\vec{a}|d)$ followed by successive marginalization provides constraints on all model parameters and a measure of the false alarm probability. The difficulty in practice is in performing the integrals over parameter space. We first consider circular orbits, which have a simple sinusoidal velocity curve, and introduce some analytic approximations that allow us to rapidly carry out these integrals. Apart from being a testing ground for these techniques which we will then apply to eccentric orbits, fitting circular orbits is actually quite useful since sinusoid fits are sufficient to detect orbits even with moderate eccentricities ($e\la 0.5$; \citealt{endl02,c04}).

For a circular orbit, the model for the velocities is 
\begin{equation}\label{eq:circmodel}
V_i=\gamma+K\sin\left(\omega t_i+\phi\right)
\end{equation}
which has four parameters: $\gamma$ is the systemic velocity, $K$ is the velocity semi-amplitude, $\phi$ the phase and $\omega=2\pi/P$ the orbital frequency, $P$ is the orbital period. Our aim in this section is to obtain $\prob(P,K)$, marginalizing over $\gamma$ and $\phi$. We first marginalize analytically over $\gamma$ to obtain $\prob(\phi,K,P|d)$, and then present two different methods for efficiently marginalizing over the parameters $K$ and $\phi$.  The methods are summarized and applied to an example data set in \S\ref{sec:circsummary}.

\subsection{Analytic marginalization of the systemic velocity}

To integrate over $\gamma$, we note again that $\chi^2$ depends quadratically on $\gamma$ around the best-fit value, as given by equation (\ref{eq:gammachi}), where this time $\gamma_0(\phi,K,P)$ is the best fit systemic velocity at each $\phi$, $P$ and $K$, that is $\gamma_0(\phi,K,P)$ is the value of $\gamma$ that minimizes $\chi^2$ at each $\phi$, $P$ and $K$, and $\chi^2(\gamma_0)$ is the corresponding minimum value of $\chi^2$. The best-fit systemic velocity can be calculated from $\partial\chi^2/\partial\gamma=0$, giving
\begin{equation}\label{eq:gamma0}
\gamma_0=\sum w_i \left[v_i-K\sin\left(\omega t_i+\phi\right)\right]/\sum w_i.
\end{equation}
Adopting a uniform prior for $\gamma$ and integrating for $\Delta\gamma\rightarrow\infty$, we find 
\begin{equation}\label{eq:probkpphi}
\prob\left(d|\phi,K,P\right)=\left(\chi^2\left[\gamma_0,\phi,K,P\right]\right)^{-(N-1)/2},
\end{equation}
where $\gamma_0(\phi,K,P)$ is given by equation (\ref{eq:gamma0}), and we have set the prefactor equal to unity as in \S 2.5.

\subsection{Evaluation of $\prob(\phi,K,P|d)$ on a grid}

Next, we describe a method for rapidly evaluating $\prob(\phi,K,P|d)$ numerically for a grid of values of $\phi$, $K$, and $P$. We introduce the averages
\begin{eqnarray}\label{eq:avs}
\langle v\rangle&=&\sum w_iv_i/\sum w_i\nonumber\\
\langle C\rangle&=&\sum w_i \cos(\omega t_i)/\sum w_i\nonumber\\
\langle S\rangle&=&\sum w_i \sin(\omega t_i)/\sum w_i\nonumber\\
\langle vC\rangle&=&\sum w_i v_i\cos(\omega t_i)/\sum w_i\nonumber\\
\langle vS\rangle&=&\sum w_i v_i\sin(\omega t_i)/\sum w_i\nonumber\\
\langle C^2\rangle&=&\sum w_i \cos^2(\omega t_i)/\sum w_i\nonumber\\
\langle S^2\rangle&=&\sum w_i \sin^2(\omega t_i)/\sum w_i\nonumber\\
\langle SC\rangle&=&\sum w_i \cos(\omega t_i)\sin(\omega t_i)/\sum w_i
\end{eqnarray}
In this notation equation (\ref{eq:gamma0}) can be written $\gamma_0=\av{v}-K\av{C}\sin\phi-K\av{S}\cos\phi$. Substituting this expression into $\chi^2$ and simplifying, we find
\begin{eqnarray}\label{eq:chi2phikp}
{\chi^2(\phi,K,P)\over \sum w_i}=\avv{v^2}-2K\left[\avv{vC}\sin\phi+\avv{vS}\cos\phi\right]\nonumber\\
+K^2\left[\avv{C^2}\sin^2\phi+\avv{S^2}\cos^2\phi\right .\nonumber\\\left .+2\avv{SC}\sin\phi\cos\phi\right],
\end{eqnarray}
where $\avv{fg}=\av{(f-\av{f})(g-\av{g})}=\av{fg}-\av{f}\av{g}$. 

Equation (\ref{eq:chi2phikp}) allows efficient calculation of $\chi^2$ for multiple values of the parameters $\phi$,$K$ and $P$. Given three vectors --- a vector of $K$ values, a vector of $\phi$ values (and corresponding values of $\sin\phi$ and $\cos\phi$), and a vector of orbital periods and the corresponding averages over the data (terms in angle brackets) --- a 3-dimensional matrix of $\chi^2$ values can be quickly generated. The advantage is that the sums over the data need to be calculated only once, rather than being reevaluated for each new choice of $K$ and $\phi$.

Marginalizing over $\phi$ is then straightforward, since the integral
\begin{eqnarray}\label{eq:grid2}
\prob(d|K,P)&=&{1\over 2\pi}\int_0^{2\pi} d\phi\ \prob(d|\phi,K,P)\nonumber\\
&=&{1\over 2\pi}\int_0^{2\pi} d\phi\ \left(\chi^2(\phi,K,P)\right)^{-(N-1)/2}
\end{eqnarray}
can be calculated using a quadrature method based on the values of $\phi$ in the grid. To calculate the odds ratio, we should compare this with the probability for a no planet model, which has $V_i=\gamma$ only. In this case, $\gamma_0=\av{v}$, and $\chi^2_0/\sum w_i=\avv{v^2}$, so that
\begin{equation}
\prob(0|d)=\left(\avv{v^2}\sum w_i\right)^{-(N-1)/2},
\end{equation}
which can be used in equation (\ref{eq:oddscirc}) for $\Lambda$.

\subsection{Analytic marginalization of $\phi$ and $K$}

The reason that we could analytically integrate over $\gamma$ is that the model $V_i$ is linear in $\gamma$. Now in fact, we can perform a similar analytic integration over $K$ and $\phi$ by rewriting equation (\ref{eq:circmodel}) in terms of the linear parameters $A$ and $B$,
\begin{equation}\label{eq:circlinear}
V_i=\gamma+A\sin\omega t_i+B\cos\omega t_i
\end{equation}
where $A=K\cos\phi$ and $B=K\sin\phi$. In seminal papers on Bayesian signal detection, \cite{bretthorst} carried out analytic integration over $A$ and $B$, and we follow the same approach here (see also \citealt{ford08}).

To perform the integration, we use the fact that the quadratic shape of $\chi^2$ that we found for $\gamma$ (eq.~[\ref{eq:gammachi}]) generalizes to an arbitrary linear model $V_i=\sum_k a_k g_k(t_i)$. It is straightforward to show that\footnote{There is an approximation known as the Laplace approximation \citep{sivia} in which the quadratic form in equation (\ref{eq:generalquadratic}) is assumed close to the mimimum $\chi^2$ value. \cite{ford08} applied this approximation to circular orbit fits at specified orbital periods, but in fact as we have noted here the approximation is exact in this case because the model is linear. We have tried applying the Laplace approximation to carry out the integral in $\phi$ in eq.~[\ref{eq:grid2}]. However, we find that this approximation does not perform well at low $K$, where $\prob(d|\phi)$ is bimodal, and in addition is not convenient numerically as it requires a search for the peak in $\prob(d|\phi)$ at each value of $K$.}\begin{equation}\label{eq:generalquadratic}
\chi^2(\vec{a})=\chi^2(\vec{a_0})+\delta\vec{a}\cdot\alpha\cdot\delta\vec{a}
\end{equation}
where the matrix $\alpha$ is the inverse of the correlation matrix \citep{nr}, and has components $\alpha_{kl}=(1/2)(\partial^2\chi^2/\partial a_k\partial a_l)=\sum w_i g_k(t_i) g_l(t_i)$. The marginalization integral with uniform priors for the parameters can be done analytically\footnote{To prove equation (\ref{eq:linearmarg}), follow the method given in the Appendix of \cite{sivia}, where a similar result is derived for a likelihood $\propto\exp(-\chi^2/2)$.}
\begin{equation}\label{eq:linearmarg}
\int d^m\vec{a}\ \left(\chi^2\right)^{-N/2}={\left(\chi^2_0\right)^{-{N-m\over 2}}\over \sqrt{{\rm det}\ \alpha}}{\pi^{m/2}\Gamma\left({N-m\over 2}\right)\over \Gamma\left({N\over 2}\right)},
\end{equation}
where $m$ is the number of parameters integrated over. We use subscript zero to indicate the best fit value of parameters, or the corresponding minimum value of $\chi^2$.

Applying this result to the integration over $A$ and $B$ gives
\begin{eqnarray}\label{eq:probP}
\prob(P|d)&=&{1\over P}\int {dAdBd\gamma\over \Delta A\Delta B\Delta\gamma} \ \left(\chi^2(A,B,\gamma,P)\right)^{-N/2}\nonumber\\
&=&{1\over P\Delta A\Delta B\Delta \gamma}{\left(\chi^2_0\right)^{-{N-3\over 2}}\over \sqrt{{\rm det}\ \alpha}}{\pi^{3/2}\Gamma\left({N-3\over 2}\right)\over \Gamma\left({N\over 2}\right)}.
\end{eqnarray}

The values of $\chi^2_0$ and $\det\alpha$ can be calculated as a function of $P$ as follows. First by minimizing $\chi^2$ with respect to $A$, $B$, and $\gamma$, the best fit values of $\gamma$, $A=K\cos\phi$ and $B=K\sin\phi$ are
\begin{eqnarray}\label{eq:c1}
A_0&=&{\avv{vS}\avv{C^2}-\avv{vC}\avv{SC}\over \avv{C^2}\avv{S^2}-\avv{SC}^2}\\
B_0&=&{\avv{vC}\avv{S^2}-\avv{vS}\avv{SC}\over \avv{C^2}\avv{S^2}-\avv{SC}^2}\\
\gamma_0&=&\av{v}-A_0\av{S}-B_0\av{C}
\end{eqnarray}
and the minimum value of $\chi^2$ is
\begin{eqnarray}
{\chi^2_0(P)\over \sum w_i}=\avv{v^2}-2A_0\avv{vS}-2B_0\avv{vC}\nonumber\\+A_0^2\avv{S^2}+B_0^2\avv{C^2}+2A_0B_0\avv{SC}
\end{eqnarray}
and
\begin{equation}\label{eq:c2}
{\det\alpha\over (\sum w_i)^3}=\avv{S^2}\avv{C^2}-\avv{SC}^2
\end{equation}
This allows us to easily calculate $\prob(P|d)$. 

The only remaining question is what to choose for the prior ranges $\Delta A$ and $\Delta B$ (the prior range in gamma $\Delta \gamma$ cancels when we form the odds ratio). Unfortunately, the analytic evaluation of the integral in equation (\ref{eq:probP}) is only possible for a uniform prior in $A$ and $B$. Since $dAdB=KdKd\phi$, a uniform prior in $A$ and $B$ corresponds to a prior $\prob(K)\propto K$ rather than the logarithmic prior $\prob(K)\propto 1/K$ that we assumed in the grid-based calculation (see discussion in \citealt{bretthorst} who chose a different prior to \citealt{jaynes}). Therefore the analytic marginalization gives more weight to large $K$ solutions, whereas the grid based approach gives more weight to small $K$ solutions. We correct for this in an approximate way by choosing the normalization appropriately. We find that the choice \begin{equation}\label{eq:DADB}\Delta A\Delta B=K_0(P) K_{0,{\rm av}}\log(K_2/K_1),\end{equation} where $K_{0,{\rm av}}$ is the best fit velocity amplitude averaged over all frequencies reproduces the normalization of the grid based calculation, with final odds ratios typically within a factor of 2. 

\subsection{The probability distribution of $K$ at each orbital period}

Analytical marginalization over the linear parameters $A$ and $B$ is convenient, but in doing so, we have thrown away information about the velocity amplitude $K$. It turns out that we can get it back very easily using an analytic approximation for the shape of $\prob(d|K)$ due to \cite{jaynes}\footnote{We follow a slightly different argument than \cite{jaynes}, but with the same spirit. The same approach was used by \cite{groth} to derive the statistical distribution of periodogram powers in the presence of a signal plus Gaussian noise, and recently \cite{shen} made a similar approximation to derive the shape of the probability density for eccentricity in a Keplerian orbit fit.}.
The idea is to assume the parameters $A$ and $B$ are uncorrelated\footnote{This is a good approximation for large $N$. The covariance between $A$ and $B$ is $\propto\sum w_i\sin\omega t_i\cos\omega t_i$ which averages to zero for large $N$.}, giving
\begin{equation}
\prob(A,B|d)\propto \exp\left[-{(A-A_0)^2\over 2\sigma_A^2}-{(B-B_0)^2\over 2\sigma_B^2}\right]
\end{equation}
where $A_0$ and $B_0$ are the best fit values, and $\sigma_A$ and $\sigma_B$ are the errors in determining $A$ and $B$ from the data. Now writing $A=K\cos\phi$ and $B=K\sin\phi$, we find
\begin{equation}
\prob(K,\phi|d)\propto  \exp\left(-{K^2\over 2\sigma_K^2}+{KK_0\over \sigma_K^2}\cos(\phi+\phi_0)\right)
\end{equation}
where $\phi_0$ is a constant that can be determined (the precise value is not important here), the best fit amplitude is $K_0=(A_0^2+B_0^2)^{1/2}$, and we assume $\sigma_A^2=\sigma_B^2=\sigma_K^2$. If the variance of the noise is $s^2$, we expect to be able to determine the amplitude $K$ to an accuracy $\sigma_K^2\approx 2s^2/N$. Using this approximation, together with the integral representation of the modified Bessel function
\begin{equation}
I_0(z)={1\over 2\pi}\int_0^{2\pi}dt\ e^{z\cos t}
\end{equation}
gives
\begin{equation}
\prob(K|d)\propto \exp\left(-{NK^2\over 4s^2}\right)I_0\left({NKK_0\over 2s^2}\right).
\end{equation}
Since we want a prior for $K$ of $\prob(K)\propto 1/K$, we divide the area element $dAdB=KdKd\phi$ by $K^2$, giving the final result
\begin{equation}\label{eq:Kpdf}
\prob(K|d)dK\propto \exp\left(-{NK^2\over 4s^2}\right)I_0\left({NKK_0\over 2s^2}\right) {dK\over K}.
\end{equation}

Comparing to the results of our grid search, we find that equation (\ref{eq:Kpdf}) reproduces the distribution of $K$ values at each orbital period remarkably well. We estimate $s^2$ as the mean square deviation of the residuals to the best fit sinusoid, or $s^2(P)=\chi^2_0(P)/\sum w_i$.
We normalize the distribution of $K$ at each $P$ so that $\int \prob(K,P|d)dK=\prob(P|d)$, where $\prob(P|d)$ is determined from the analytic marginalization over $A$ and $B$ (eq.~[\ref{eq:probP}]). This choice of normalization as a function of $P$ gives the best agreement with the grid code.

\begin{figure}
\includegraphics[width=\columnwidth]{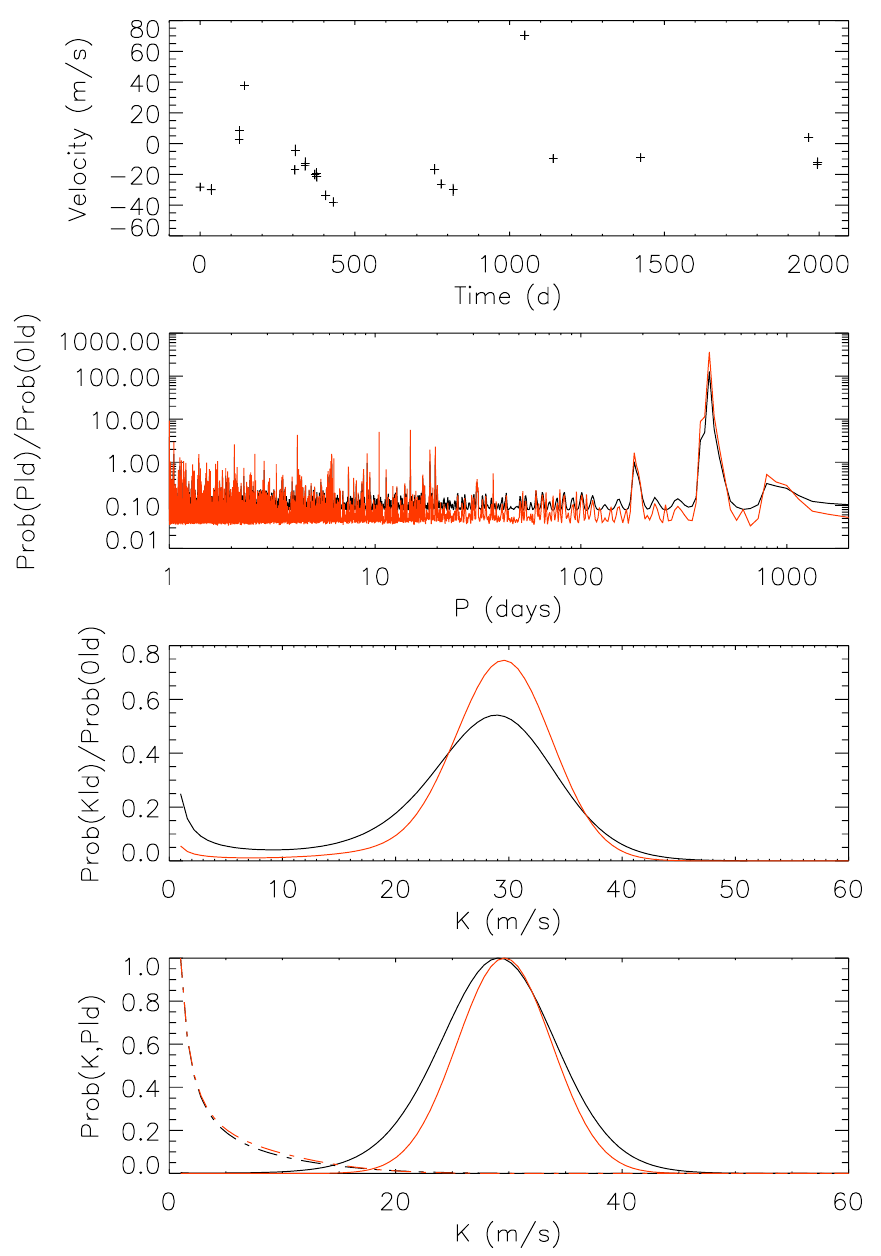}
\caption{Results of circular orbit fitting to data for HD~4203, using analytic marginalization over $K$ and $\phi$, and reconstructing $\prob(K)$ using equation (\ref{eq:Kpdf}) (red curves) and by calculating $\prob(\phi,K,P)$ on a grid (black curves). $\prob(K)$ in the third panel is normalized such that each curve has the same area beneath the curve. The bottom panel compares the $\prob(K)$ obtained for periods $420.1$ days (solid curves, close to the best-fitting frequency) and $19.0$ days (dot-dashed curves, no significant fit at this frequency) for the analytic and grid-based approaches. \label{fig:4203circ}}
\end{figure}

\subsection{Summary and example}
\label{sec:circsummary}

Let's summarize the main results of this section. We have discussed two methods for evaluating $\prob(P,K|d)$ for circular orbits. First, equations (\ref{eq:chi2phikp}) and (\ref{eq:grid2}) can be used to calculate $\chi^2(P,K,\phi)$ for many different values of $P$, $K$, and $\phi$, and from there $\prob(P,K|d)$ obtained by integration over $\phi$. No approximations are made in this approach, which we refer to as the ``grid-based approach''. Second, equations (\ref{eq:probP}) to (\ref{eq:DADB}) provide a method for evaluating $\prob(P|d)$ using analytic marginalization over the linear parameters $A$ and $B$ and therefore $K$ and $\phi$. The analytic marginalization requires that we assume a prior $\prob(K)\propto K$ rather than $1/K$, but by choosing the normalization appropriately (eq.~[\ref{eq:DADB}]), we approximately recover the results corresponding to $\prob(K)\propto 1/K$. Next, given the best fit amplitude $K_0=(A_0^2+B_0^2)^{1/2}$ at each period, $\prob(P,K|d)$ can be calculated for a grid of $K$ values using the analytic approximation of equation (\ref{eq:Kpdf}). We refer to this second approach as the ``analytic approach''.

As an example, we consider the 23 radial velocities for the star HD~4203 made available in the \cite{butlercatalog} catalog of nearby exoplanets (see \citealt{vogt02} for the original discovery of this planet). The orbital parameters given by \cite{butlercatalog} are $P=431.88\pm 0.85$ days, $K=60.3\pm 2.2\ {\rm m/s}$, and $e=0.519\pm 0.027$. They also include a linear long term trend of $-4.38\pm 0.71\ {\rm m\ s^{-1}\ yr^{-1}}$. The rms of the residuals to this solution is $4.1\ {\rm m/s}$.

We use both of the techniques described above to fit a circular orbit plus constant to this data. We consider orbital periods between 1 day and the time-span of the data, $T=2000$ days. We evaluate $4 N_f$ frequencies, where $N_f=(\Delta f)T$ is the estimated number of independent frequencies in the frequency range $\Delta f$ \citep{c04}. The values of $K$ considered range from $1\ {\rm m/s}$ to $2\Delta v$, where $\Delta v$ is the range of the measured velocities. For the grid-based approach, we find that the typical time required on a 2--3 GHz CPU is $\sim 10^{-7}\ {\rm s}$ per set of parameters $(\phi,K,P)$, so that for example 3000 periods, 100 values of $K$ and $30$ phases, or $10^7$ total combinations, takes $1\ {\rm s}$ to evaluate. For $\phi$, we align the grid with the best fit phase $\phi_0$ at each $P$. In this way, we guarantee that the best fit value of $\phi$ is included on the grid, which reduces the number of grid points we need to use in $\phi$. The analytic marginalization technique requires $\sim 5\times 10^{-7}\ {\rm s}$ per $P$ and $K$ value, so that a search of 3000 periods, keeping track of 100 values of $K$ takes $\sim 0.1\ {\rm s}$. We use the routine bessi0 from \cite{nr} to calculate the Bessel function in equation (\ref{eq:Kpdf}).

Figure \ref{fig:4203circ} compares the two techniques. The red curves show the results of the analytic marginalization, the black curves show the results of the grid-based calculation. The false alarm probabilities are $0.14$ (grid) and $0.060$ (analytic) (odds ratios $6.3$ and $16$ respectively). The distribution of $K$ agrees well between the two techniques, although the probability curve is shifted to larger values of $K$ for the analytic approach compared to the grid approach, consistent with the different priors. The false alarm probability $\sim 0.1$ means that this would not count as a detection. This is an example of a case in which the large eccentricity $e>0.5$ prevents detection by fitting circular orbits. The best fit amplitude $K\approx 30\ {\rm m/s}$ for circular orbits is significantly smaller than for the Keplerian orbit fit of \cite{butlercatalog}. Using 100 values of $K$ between $1\ {\rm m/s}$ and $60\ {\rm m/s}$, we find the 99\% upper limit on $K$ is $41.2\ {\rm m/s}$ (analytic) or $41.3\ {\rm m/s}$ (grid). The bottom panel in Figure \ref{fig:4203circ} compares the probability distribution of $K$ at two different periods obtained from the grid-based approach and the analytic approach. This shows that equation (\ref{eq:Kpdf}) reproduces the distribution from the grid-based calculation well.


\section{Eccentric orbits}

We now consider full Keplerian fits to the data. The techniques we developed in the previous section for circular orbits can be readily applied to Keplerian orbits, because the Keplerian model is linear in a subset of parameters which can therefore be treated analytically, as we now describe.

\subsection{Calculation of $\prob(P,K,e|d)$}

For a Keplerian orbit, the radial velocity can be written
\begin{equation}
V=\gamma+K\left[\cos(\theta+\omega_p)+e\cos\omega_p\right]
\end{equation}
where $K$ is the velocity amplitude, $e$ is the eccentricity of the orbit, $\omega_p$ is the argument of periastron\footnote{We write it as $\omega_p$ to distinguish it from the orbital frequency $\omega=2\pi/P$.}. The true anomaly $\theta$ is a function of the time $t$ and the three parameters $e$, $P$, and $t_p$, where $t_p$ is the time of periastron passage (acting as an overall phase for $V(t)$). To calculate $\theta(t; e,P,t_p)$, we must solve the relations
\begin{equation}\label{eq:theta}
\tan\left({\theta\over 2}\right)=\left({1+e\over 1-e}\right)^{1/2}\tan\left({E\over 2}\right)
\end{equation}
\begin{equation}\label{eq:kepler}
E-e\sin E=M={2\pi\over P}(t-t_P)
\end{equation}
where $E$ is the eccentric anomaly, and $M$ the mean anomaly.

The first point to note is that the six orbital parameters, $\vec{a}=(\gamma,K,\omega_p,P,e,t_p)$ can be divided into two groups, ``slow'' and ``fast'' parameters, $\vec{a_s}=(P,e,t_p)$ and $\vec{a_f}=(\gamma,K,\omega_p)$ respectively. Each time we change a value of the slow parameters, we must re-solve equations (\ref{eq:theta}) and (\ref{eq:kepler}) to calculate the values of $\theta$, whereas when we change a value of the fast parameters only we do not need to recalculate the values of $\theta$. This is reminiscent of the division into fast and slow parameters in analysis of CMB data (e.g.~\citealt{lewisbridle02,tegmark04}). We can use this division to increase the speed of the parameter search.

For a given set of the slow parameters, we can find the best fitting fast parameters with a linear least-squares fit, since we can write 
\begin{equation}\label{eq:ecclinear}
V=A\sin\theta+B\cos\theta+\tilde{\gamma}
\end{equation}
with $A=-K\sin\omega_p$, $B=K\cos\omega_p$, and $\tilde{\gamma}=\gamma+Ke\cos\omega_p$. A linear least-squares fit returns the best-fitting values of $A$,$B$, and $\tilde{\gamma}$, and therefore $K$ ($K^2=A^2+B^2$), $\omega_p$ ($\tan\omega_p=-B/A$), and $\gamma$. This halves the number of parameters that we need to search to find the best-fitting solution.

The fact that the fast parameters $\vec{a_f}$ can be obtained from a linear fit means that we can directly apply the techniques we developed for circular orbits in \S 3 to marginalize over them. For the grid-based approach, equation (\ref{eq:chi2phikp}) should be replaced by 
\begin{eqnarray}
{\chi^2(K,\omega_p)\over \sum w_i}=\avv{v^2}+2K\left[\avv{vS}\sin\omega_p-\avv{vC}\cos\omega_p\right]\nonumber\\+K^2\left[\avv{C^2}\cos^2\omega_p+\avv{S^2}\sin^2\omega_p\right .\nonumber\\\left .-\avv{SC}2\sin\omega_p\cos\omega_p\right]
\end{eqnarray}
where $\omega_p$ now plays the same role as $\phi$ for circular orbits, and the sums over the data involve $\theta_i$ rather than $\omega t_i$. For example, the definition of $\av{S}$ in equation (\ref{eq:avs}) should be replaced by $\av{S}=\sum w_i\sin\theta_i/\sum w_i$. 

Similarly, since equations (\ref{eq:circlinear}) and (\ref{eq:ecclinear}) are of the same form, the analytic integration over $A$ and $B$ can be applied directly to the Keplerian case, giving $\prob(P, e, t_p|d)$ analytically from equations (\ref{eq:probP}) to (\ref{eq:DADB}). As for circular orbits, the distribution of velocity amplitude at each $(P,e,t_p)$, $\prob(K,P,e,t_p|d)$, can be recovered, being well-approximated by equation (\ref{eq:Kpdf}).

\begin{figure}
\includegraphics[width=\columnwidth]{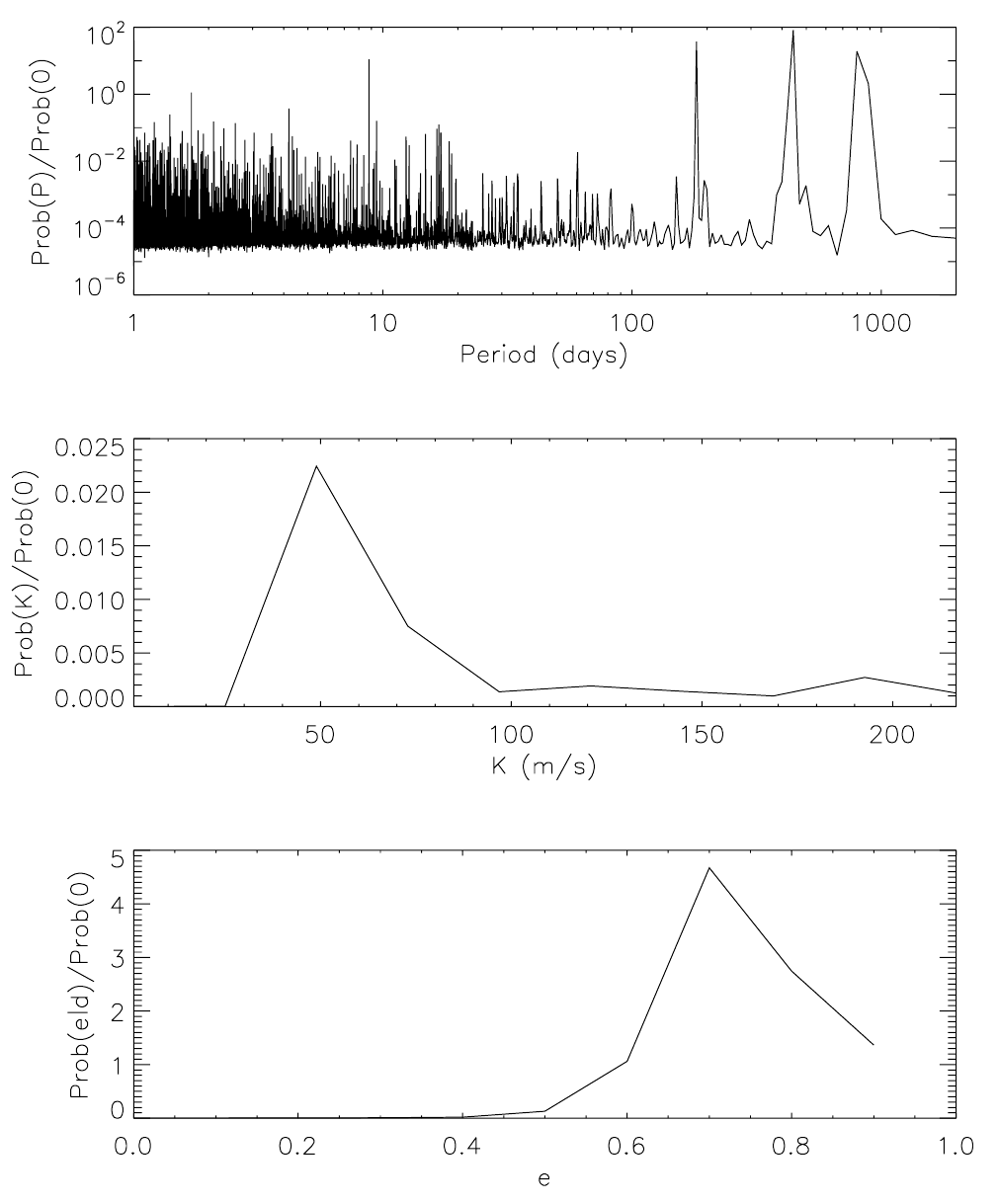}
\caption{Results of Keplerian fits to the HD 4203 data from \citealt{butlercatalog}, including a linear trend. In this coarse scan of parameter space, $\prob(P,e,K|d)$ is calculated for 10 eccentricities between 0 and 0.9, 10 velocities between 1 and $217\ {\rm m/s}$ (twice the velocity span of the data), and 7978 periods between 1 day and 1996 days (the time span of the data).
\label{fig:4203ecccoarse}}
\end{figure}

\begin{figure*}
\includegraphics[width=7in]{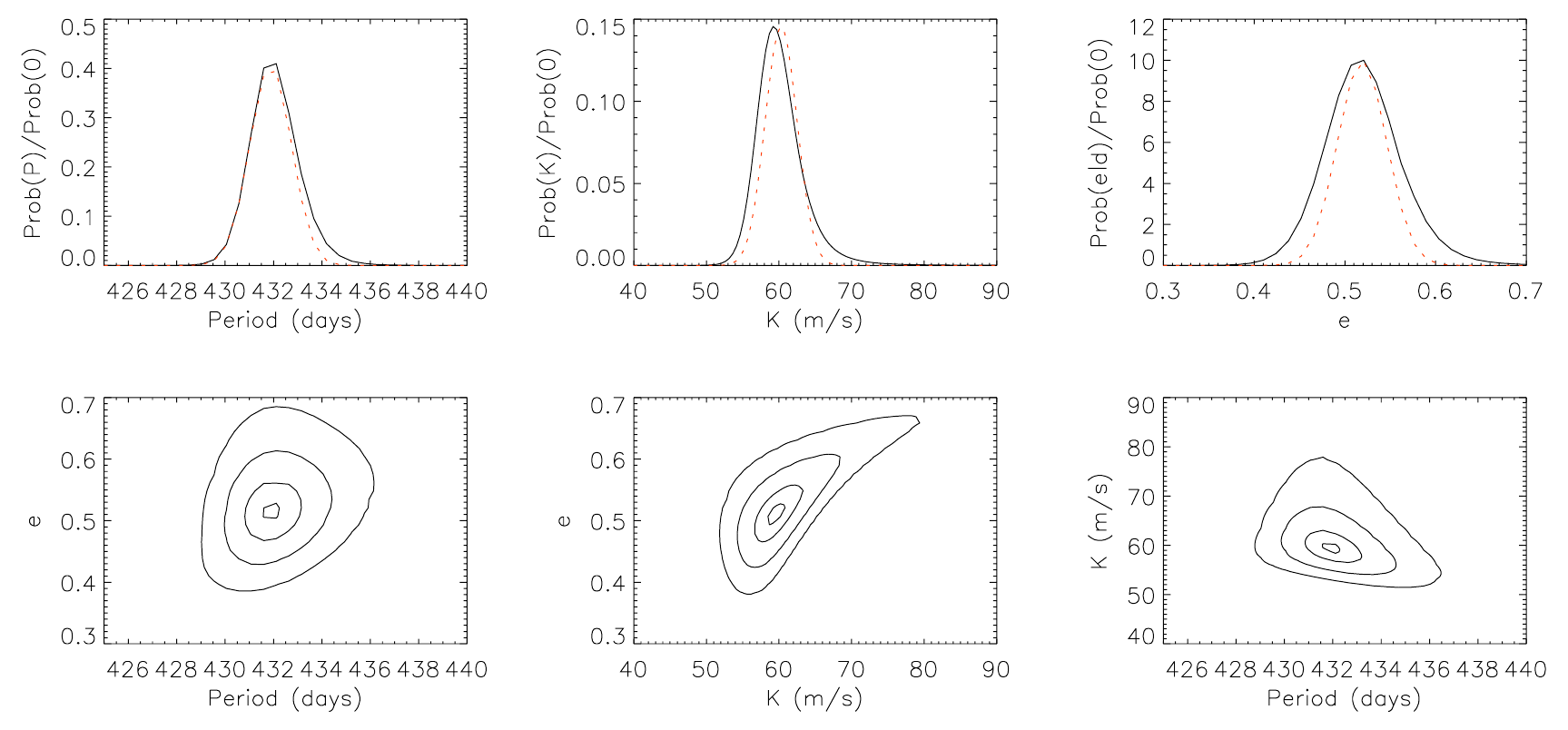}
\caption{Results of Keplerian fits to the HD 4203 data from \citealt{butlercatalog}, including a long term trend. The dotted curves show Gaussian distributions with central values and standard deviations matching those given by \citealt{butlercatalog}. 100 values of K, 30 eccentricities and 30 periods were calculated in the range shown. The contours enclose 10\%, 50\%, 90\% and 99\% of the probability.
\label{fig:4203ecc}}
\end{figure*}

\subsection{Example}

As an example, we return to the HD~4203 data considered previously. We first calculate $\prob(P,e|d)$ for a grid in $P$ and $e$. The integration over $t_p$ is carried out using a simple algorithm in which we double the number of equally-spaced $t_p$ values until the required accuracy is obtained. For each combination of $P$, $e$, and $t_p$ considered, we analytically integrate over $\gamma$, $K$, and $\omega_p$, and at the same time use equation (\ref{eq:Kpdf}) to keep track of $\prob(K; P,e,t_p|d)$. We use Newton's method to solve Kepler's equation, taking advantage of the fact that the required derivative can be calculated analytically. Our implementation of this algorithm takes $\approx 5\times 10^{-5}\ {\rm s}$ per $P$, $e$, and $t_p$ value considered, with 30 $K$ values tracked through the calculation. For an average 200 values of $t_p$, 10 eccentricities, and 3000 periods, the total time needed is $\approx 30\ {\rm s}$ for a scan of parameter space. We have also implemented the grid-based approach, and find that it is about 10 times slower than the analytic approach. The results agree well between both techniques.

The results for HD~4203 are shown in Figures \ref{fig:4203ecccoarse} and \ref{fig:4203ecc}. We first run a coarse scan of the parameter space for a single Keplerian orbit plus a linear trend. We calculate $4N_f\approx 8000$ frequencies, corresponding to the period range $1$ day to $\approx 2000$ days (the time span of the data), 10 eccentricities between 0 and 0.9, 10 velocities between $1\ {\rm m/s}$ and $216\ {\rm m/s}$ (twice the velocity span of the data). The resulting constraints on $P$, $e$ and $K$ are shown in Figure \ref{fig:4203ecccoarse}. The odds ratio is $4\times 10^4$ for the Keplerian orbit plus linear trend compared to a constant velocity model. We show the results including a linear trend, because the best fit model presented by \cite{butlercatalog} includes a trend, but in fact our results at this stage do not require a trend. The odds ratio for a similar search but without the linear term is $5\times 10^4$.

We then carry out a more detailed calculation of the parameter space near the best fitting model corresponding to the peak in $\prob(P|d)$ at $\approx 440$ days in Figure \ref{fig:4203ecccoarse}. The results are shown in Figure \ref{fig:4203ecc}. The odds ratio is $9.5\times 10^{10}$ for a Keplerian orbit plus trend compared to a constant model (for the ranges of parameters shown in Fig.~\ref{fig:4203ecc}). The much larger value of the odds ratio compared to our coarse calculation is because the parameter space considered is smaller and the peak in $\prob(P,e,K|d)$ has now been resolved. We can renormalize the odds ratio to correspond to the full range of parameter space considered in the coarse search by multiplying by the ratio of $\log P_2/P_1$ and $\log K_2/K_1$ in each calculation. Doing this, we find an odds ratio $7.4\times 10^7$. Without the linear trend the odds ratio is 100 times smaller, $7\times 10^5$, normalized to the full range of parameters. This indicates that a model with a linear trend is strongly preferred given this data. Without the linear trend, the probability peaks at similar values of $P$ and $K$, but with a larger eccentricity, $e\approx 0.7$.

The dotted curves in Figure \ref{fig:4203ecc} show Gaussian distributions with the central values and standard deviations given by \cite{butlercatalog} for $K$, $P$, and $e$. Overall there is good agreement with the central values and widths.

Repeating the calculation shown in Figure \ref{fig:4203ecc} with the grid-based method for marginalizing over $K$ and $\omega_p$ gives almost identical constraints on orbital parameters, but a smaller odds ratio by a factor of two, $3.5\times 10^7$ compared to $7.4\times 10^7$. We have also checked that other peaks in $\prob(P|d)$ that can be seen in Figure \ref{fig:4203ecccoarse} do not contribute significantly to the odds ratio. The next most important is the peak at $P\approx 800$ days, but its odds ratio is 400 times smaller than the peak at 432 days shown in detail in Figure \ref{fig:4203ecc}.

The coarse sampling for HD~4203 gave an odds ratio that was a factor of 400 smaller than the final odds ratio obtained by zooming in on the most significant peak. We find that increasing the period sampling by a factor of two to $8T\Delta f$ gives an odds ratio from the coarse search in good agreement with the odds ratio from zooming in on the peak.


\section{Comparison with previous work} 

In \S 4, we presented an algorithm that can efficiently compute $\prob(P,K,e|d)$ for a radial velocity data set. As described in \S \ref{sec:overview}, this contains information about the constraints on $P$, $K$, and $e$ and also allows a false alarm probability to be calculated. We now use our algorithm to recalculate results in the literature from MCMC and other techniques and compare.

\begin{figure}
\includegraphics[width=\columnwidth]{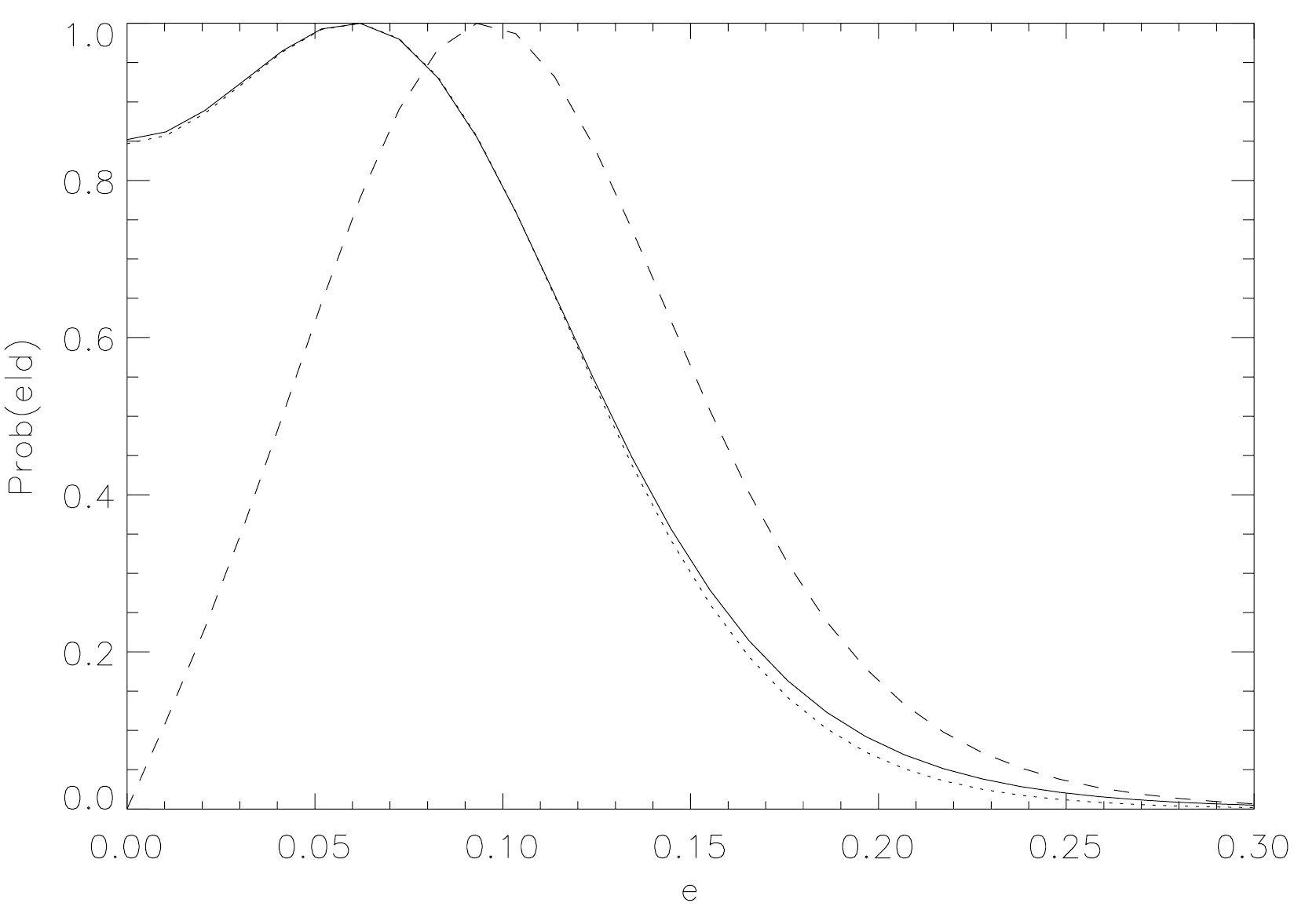}
\caption{The eccentricity distribution derived for HD 76700, using data from \citealt{tinney03}. The solid curve is for analytical marginalization over the noise scaling parameter $k$, the dotted curve is for $k=1$, and the dashed curve shows $e\prob(e|d)$, corresponding to a uniform prior in $d(e\cos\omega_p)d(e\sin\omega_p)$.
\label{fig:76700}}
\end{figure}

\subsection{Orbital parameter constraints from MCMC calculations}

\cite{ford05} used a MCMC calculation to study the constraints on orbital parameters from radial velocity data, and this paper has been followed by several others \citep{ford06,ford08,fordgregory07,gregory05,gregory07a,gregory07b,balan}. We have calculated the constraints on orbital parameters for the different single planet cases considered in these papers, and overall the agreement is excellent.

One difference is that in several published cases, the posterior probability for eccentricity drops towards zero at low eccentricities, whereas we find $\prob(e|d)$ is approximately constant as $e$ goes to zero. For HD~76700, this difference appears to be because of the different prior assumed by \cite{ford05}. The MCMC calculations in that paper take steps in $e\cos\omega_p$ and $e\sin\omega_p$ in such a way that the assumed prior is uniform in $d(e\cos\omega_p)d(e\sin\omega_p)$ giving a prior $e\,de\,d\omega_p\propto e$. In Figure \ref{fig:76700}, we allow for this different prior by plotting $e\prob(e|d)$, and the result compares favorably with Figure 2 of \cite{ford05}. (\citealt{ford05} discusses the use of importance sampling, in which the samples are weighted $\propto 1/e$ to give an effective prior uniform in $e$, but this does not seem to have been applied in Figure 2 of that paper).

\begin{figure}
\includegraphics[width=\columnwidth]{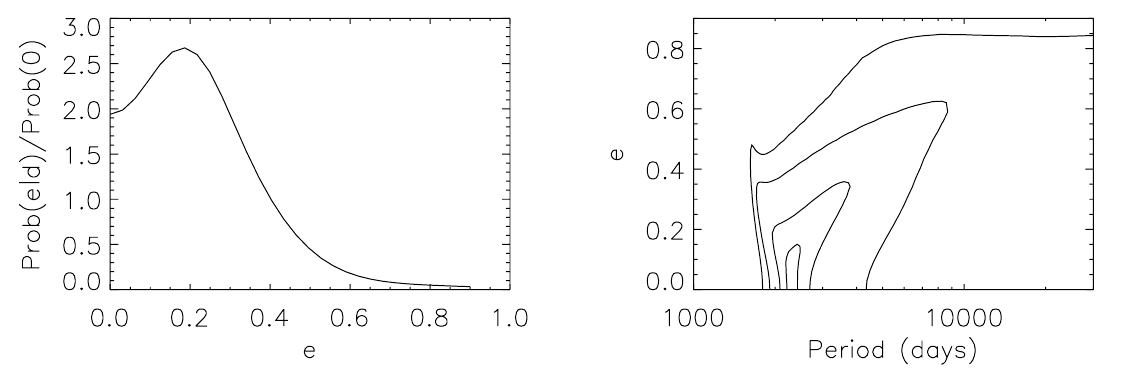}
\includegraphics[width=\columnwidth]{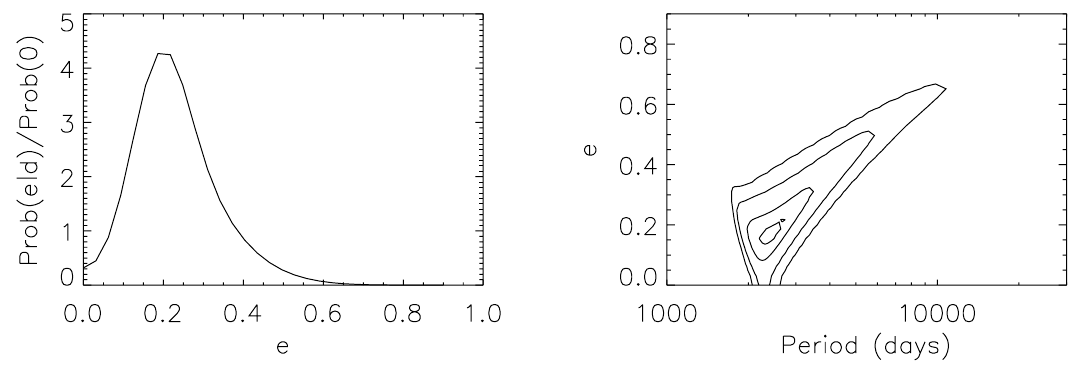}
\caption{The eccentricity distribution and joint eccentricity-orbital period distribution derived for HD 72659. The top panels are for analytic marginalization over the noise parameter $k$, whereas the bottom panels take $k=1$ as in \citealt{ford05}. Contours enclose 10, 50, 90 and 99\% of the probability.
\label{fig:72659}}
\end{figure}

For HD~72659, marginalization over the extra noise source opens up considerable parameter space at low eccentricity. In Figure \ref{fig:72659}, we show the constraints on eccentricity and period with $k$ fixed at $k=1$ and with $k$ marginalized over. \cite{ford05}, unlike later papers (e.g.~\citealt{ford06}) does not include an additional noise term, and our results for $k=1$ compare well with Figures 4 and 5 of that paper.

\subsection{Odds ratios from Gregory's parallel tempering MCMC approach}

In a series of papers, Gregory has developed a MCMC code which uses parallel tempering to exchange information between chains running with different ``temperatures''. Combining the results of different chains gives the total posterior probability for the model, allowing calculations of odds ratios and therefore model comparisons. 

\cite{gregory05} analyzed 18 radial velocities for HD~73526 from \cite{tinney03}. The period range was from 0.5 days to 3732 days, and velocities from 0 to $400\ {\rm m/s}$ using a Jeffrey's prior with a break at $1\ {\rm m/s}$. An additional noise term was added which was allowed to range between $0$ and $100\ {\rm m/s}$. He pointed out that there were two additional possible solutions with $P\approx 128$ and $376$ days besides the previously obtained solution at $P\approx 191$ days. A chain covering the entire parameter space did not converge, and so separate chains were run focussing on each of the three probability peaks. The odds ratio for a planet compared to a constant velocity was found to be $9.3\times 10^5$ (Table 5 of \citealt{gregory05}). We ran a calculation with the same period and velocity range as \cite{gregory05} (except that we take the lower bound in $K$ to be $1\ {\rm m/s}$ with a Jeffrey's prior) ($9940$ frequencies, 30 eccentricities and 30 velocities).  The odds ratio was $3.3\times 10^6$. Zooming in on the three peaks gives probability distributions for $e$, $P$, and $K$ that are very similar to the results of \cite{gregory05}. The odds ratios for the $P\approx 128$, $190$, and $376$ day peaks are $2.4\times 10^4$, $1.1\times 10^5$, and $1.0\times 10^6$ (assuming the full prior range so that these numbers can be compared). The sum of these, $1.1\times 10^6$ agrees well with the odds ratio found by \cite{gregory05} whose odds ratio includes only these three peaks. The relative probabilities of the three peaks are 2\%, 10\% and 88\%.  \cite{gregory05} found relative probabilities of 4\%, 3\% and 93\%.

\cite{gregory07a} found evidence for a second planet in HD~208487; we compare to their odds ratio and posterior probability for a one-planet fit. The posterior probability distributions were calculated for the 35 velocities from \cite{butlercatalog}. We find excellent agreement with the distributions of $P$, $e$ and $K$ shown in Figure 7 of \cite{gregory07a}. The odds ratio for a single planet model for this data (Table 6 of \citealt{gregory07a}) was $1.7$--$2.6\times 10^4$ for two different choices of the turnover in the modified Jeffrey's prior for the extra noise scale. For the parameter ranges in Figure 7 of \cite{gregory07a}, we find an odds ratio of $1.4\times 10^8$. Rescaling to a velocity range $1$--$2129\ {\rm m/s}$, and period range $1$ day to 1000 years, this becomes $6.1\times 10^4$, a factor of 3 times greater than \cite{gregory07a}. (The details of the priors were different, for example, the upper limit on velocity in \citealt{gregory07a}'s prior depended on period and eccentricity, but we expect this to give only a small difference).

\cite{gregory07b} presented evidence for three planets in HD~11964 from 87 radial velocities in the \cite{butlercatalog} catalog. The odds ratio reported for the single planet model is $3\times 10^9$ (Table 4 of \citealt{gregory07b}). We find an odds ratio in good agreement, $2\times 10^9$. Although \cite{gregory07b} does not show posterior probability distributions for orbital parameters for the one planet model, the distributions of $P$, $e$ and $K$ we find compare well with those for the $P\approx 2000$ day signal in the three planet model of \cite{gregory07b}. For this data, \cite{butlercatalog} include a linear term. We find the odds ratio for a linear versus constant no-planet model to be 1300. Including a  linear term in the planet model gives an odds ratio of $3\times 10^6$, much smaller than the odds ratio for a planet model only. Therefore, we find that a single planet model with $P\approx 2000$ is preferred over a linear trend only or planet plus linear trend by a large factor (in agreement with \citealt{wright07} who also concluded that the trend reported by \citealt{butlercatalog} was likely spurious).

\subsection{False alarm probabilities}

\cite{marcy05} discuss the calculation of false alarm probabilities using a scrambled velocity method in which the residuals to the best-fitting Keplerian orbit are used as an estimate of the noise distribution. In that paper, they announced five new planets from the Keck Planet Search. False alarm probabilities were calculated for two cases that looked marginal, HD~45350 (FAP$<0.1$\% scrambled, $4\times 10^{-5}$ F-test) and HD~99492 (FAP$\approx 0.1$\% scrambled, $3\times 10^{-4}$ F-test). For HD~99492, we find odds ratios scaled to 1.0 for no planet are 0.33 for a linear  trend but no planet, 1.66 for a planet, 200.0 for a planet plus linear trend. Therefore, a linear trend is preferred in this case. The FAP using equation (\ref{eq:oddswithtrend}) for the odds ratio is $7\times 10^{-3}$. For HD~45350, we find odds ratios: 1.0, 0.18, $6.7\times 10^5$, $4.7\times 10^5$, giving FAP$\approx 10^{-6}$. As \cite{marcy05} noted, the evidence for a linear trend in this source is marginal (the odds ratios are similar with and without a trend).

\begin{figure}
\includegraphics[width=\columnwidth]{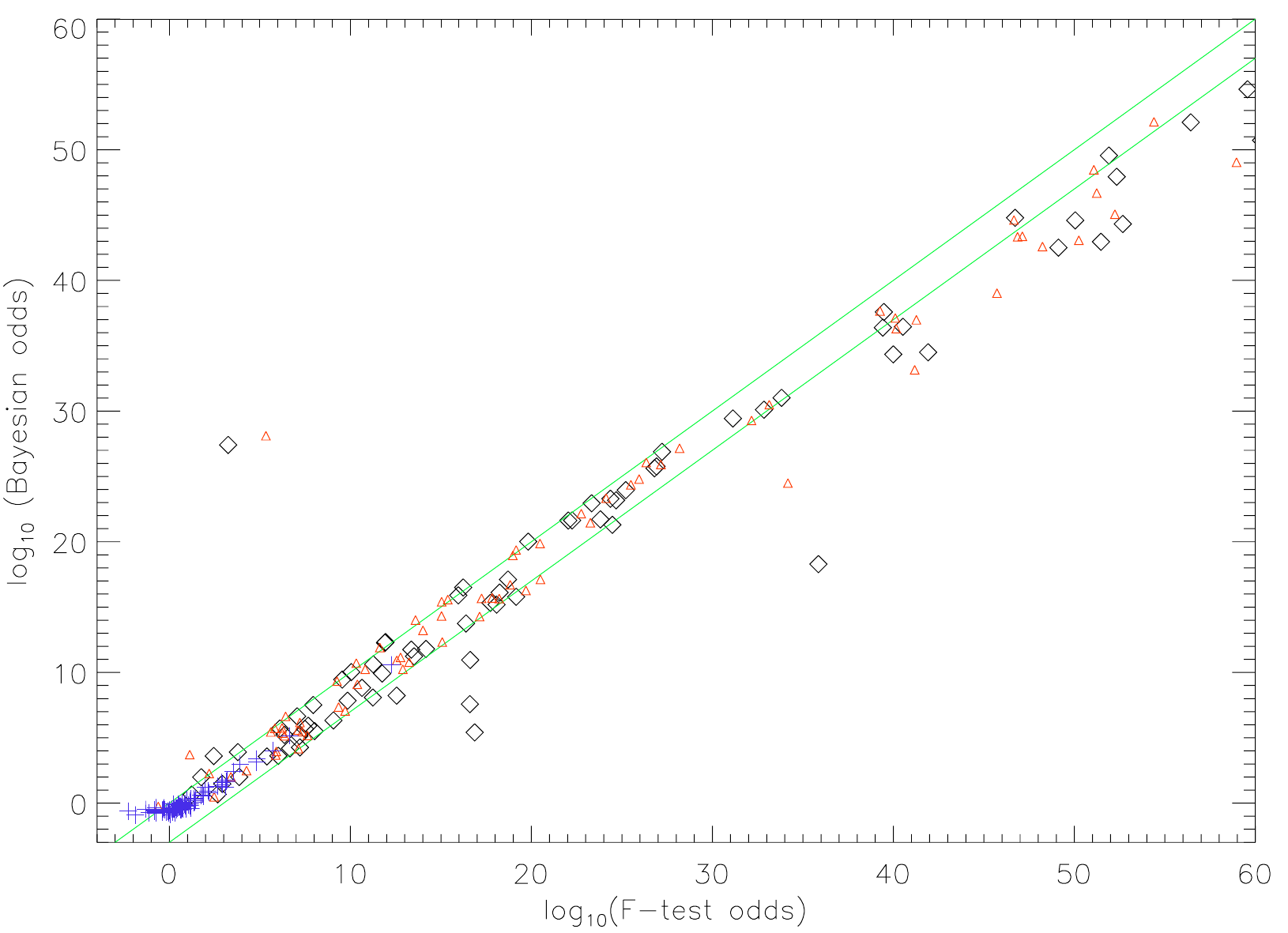}
\includegraphics[width=\columnwidth]{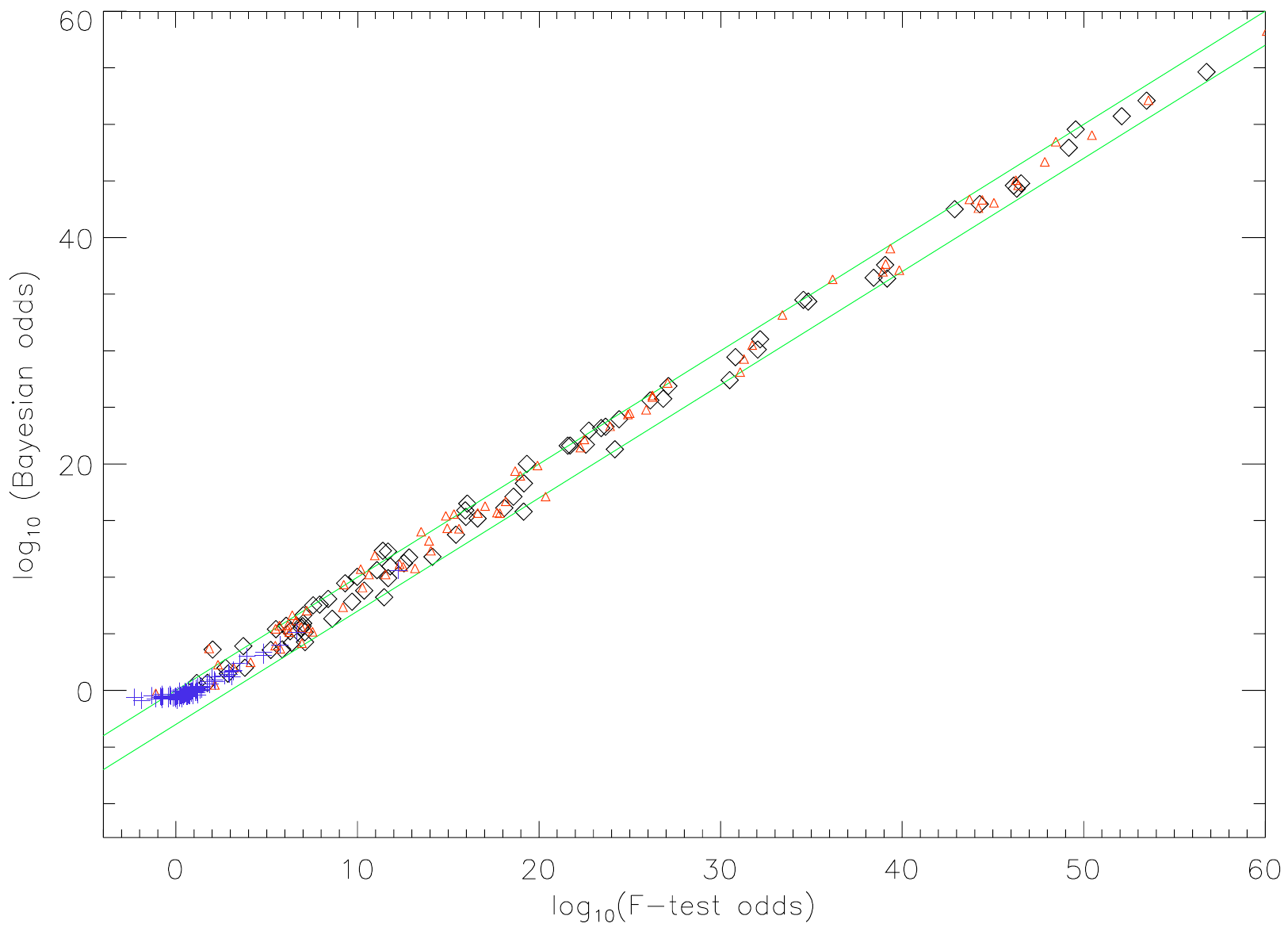}
\caption{Comparison between logarithm of the odds ratio $\log_{10}\Lambda$ from the Bayesian calculation and the F-test. The Bayesian odds ratios are from coarse sampling ($8T\Delta f$ periods, 10 eccentricities) of the 84 radial velocity data sets from Keck, Lick, and AAT published as part of the \citealt{butlercatalog} catalog. We compare with analytic F-test FAPs, converted to odds ratios using the relation $\Lambda=(1/F)-1$. The crosses are the odds ratios for a linear trend versus constant velocity, the diamonds are odds ratios for a planet versus constant, and the triangles are for planet plus long term trend versus long term trend only. The upper panel uses a Keplerian fitting routine to determine the F-test FAP, whereas in the lower panel we use the minimum $\chi^2$ found in the Bayesian routine to calculate the analytic FAP.
\label{fig:fap}}
\end{figure}

\cite{c04} described a quick estimate of the FAP based on an F-test at each independent frequency. Generalizing the Lomb-Scargle periodogram to eccentric orbits, the idea is to define a power at each frequency
\begin{equation}
z={(N-5)\Delta\chi^2\over 4\chi^2_{\rm Kep}}={(N-5)(\chi^2_0-\chi^2_{\rm Kep})\over 4\chi^2_{\rm Kep}}.
\end{equation}
For Gaussian noise, $z$ follows the $F_{4,N-5}$ distribution\footnote{Assuming that the no planet model being compared to is a constant velocity model. If a linear trend is included in the no planet model and the planet model, $z$ is defined with a factor of $N-6$ replacing $N-5$, and then follows the $F_{4,N-6}$ distribution.}, which allows a calculation of ${\rm Prob}(z>z_{\rm max})$ for an observed maximum power $z_{\rm max}$. The FAP is then 
\begin{equation}
{\rm FAP}=1-(1-{\rm Prob}(z>z_{\rm max}))^{N_f}\approx N_f{\rm Prob}(z>z_{\rm max}).
\end{equation}
The number of independent frequencies $N_f$ can be estimated as $N_f\approx T\Delta f$.

We have used this approach to calculate the FAP for the 84 stars with published radial velocities as part of the \cite{butlercatalog} catalog of exoplanets. To find $\chi^2_{\rm Kep}$, we follow the automated procedure used by \cite{c08}, which involves using the top two well-separated peaks in the Lomb-Scargle periodogram as starting periods for full Keplerian fits. To compare with the Bayesian odds ratios, we convert the F-test FAP into an odds ratio by inverting equation (\ref{eq:F}). To find Bayesian odds ratios, we run a coarse sampling of the parameter space with $8T\Delta f$ periods for each of these 84 stars, with and without a long term linear trend.  

The results are shown in Figure \ref{fig:fap}. In the lower panel, we use the minimum value of $\chi^2$ found in the Bayesian calculation to calculate the F-test FAP. In this case, the odds ratios are well-correlated, although with the Bayesian odds ratio between 1 and 1000 times smaller than the F-test odds ratio. In the upper panel, there is more scatter. This arises from differences between the minimum $\chi^2$ values found by the Keplerian fitting routine and the Bayesian routine. For example, the two points above and to the left of the upper panel of Figure \ref{fig:fap} are for HD 80606, which has a very eccentric orbit. Our Keplerian fitting routine, which uses circular orbit fits as its starting point failed to find the best-fitting solution, whereas the Bayesian routine, with its systematic scan of parameter space did find it. Generally the scatter is downwards, indicating that the Bayesian routine sometimes find a larger minimum $\chi^2$ than the Keplerian fitting routine. Likely this is due to the finite period sampling, whereas the Keplerian fitting routine can adjust the period to lower $\chi^2$.

The fact that the Bayesian odds ratios tend to be lower than the F-test odds ratios indicates that the Bayesian calculation is more conservative than the F-test. In fact, this is expected. \cite{c04} showed that the Bayesian odds ratio is closely related to the F-test (periodogram), but with a different definition for the number of independent frequencies. In the Bayesian calculation, the number of trials counts the frequencies, but also the range of the other parameters \citep{c04}. In this way, the Bayesian calculation penalizes models with larger ranges of parameters, for all parameters, not just frequency.

\begin{figure*}
\includegraphics[width=7in]{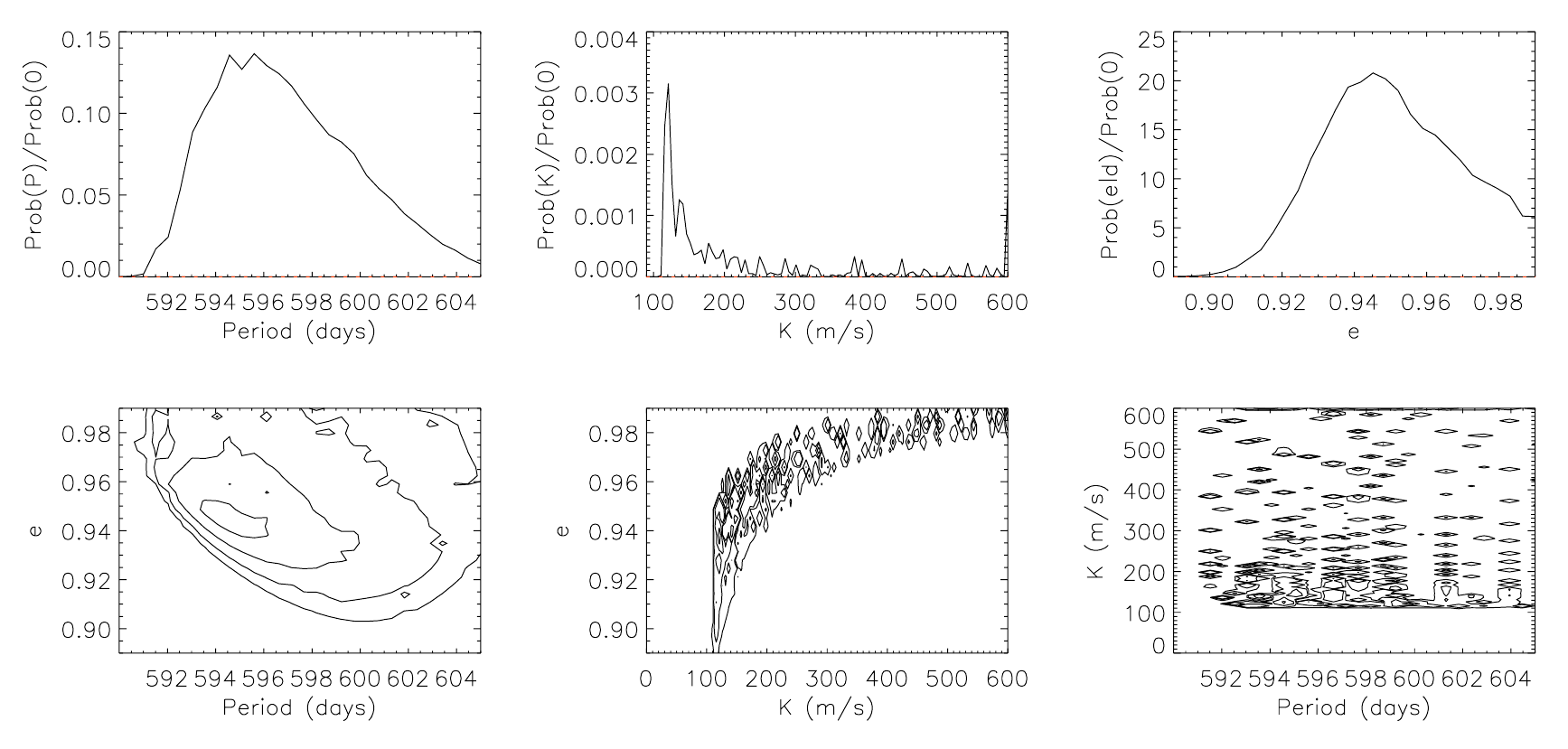}
\caption{Results of Keplerian fits to the HD~20782 data from \citealt{otoole08}. 30 eccentricities, 30 periods and 100 $K$ values were calculated in the ranges shown. Contours enclose 10\%, 50\%, 90\%, and 99\% of the total probability.\label{fig:20782}}
\end{figure*}

\begin{figure*}
\includegraphics[width=7in]{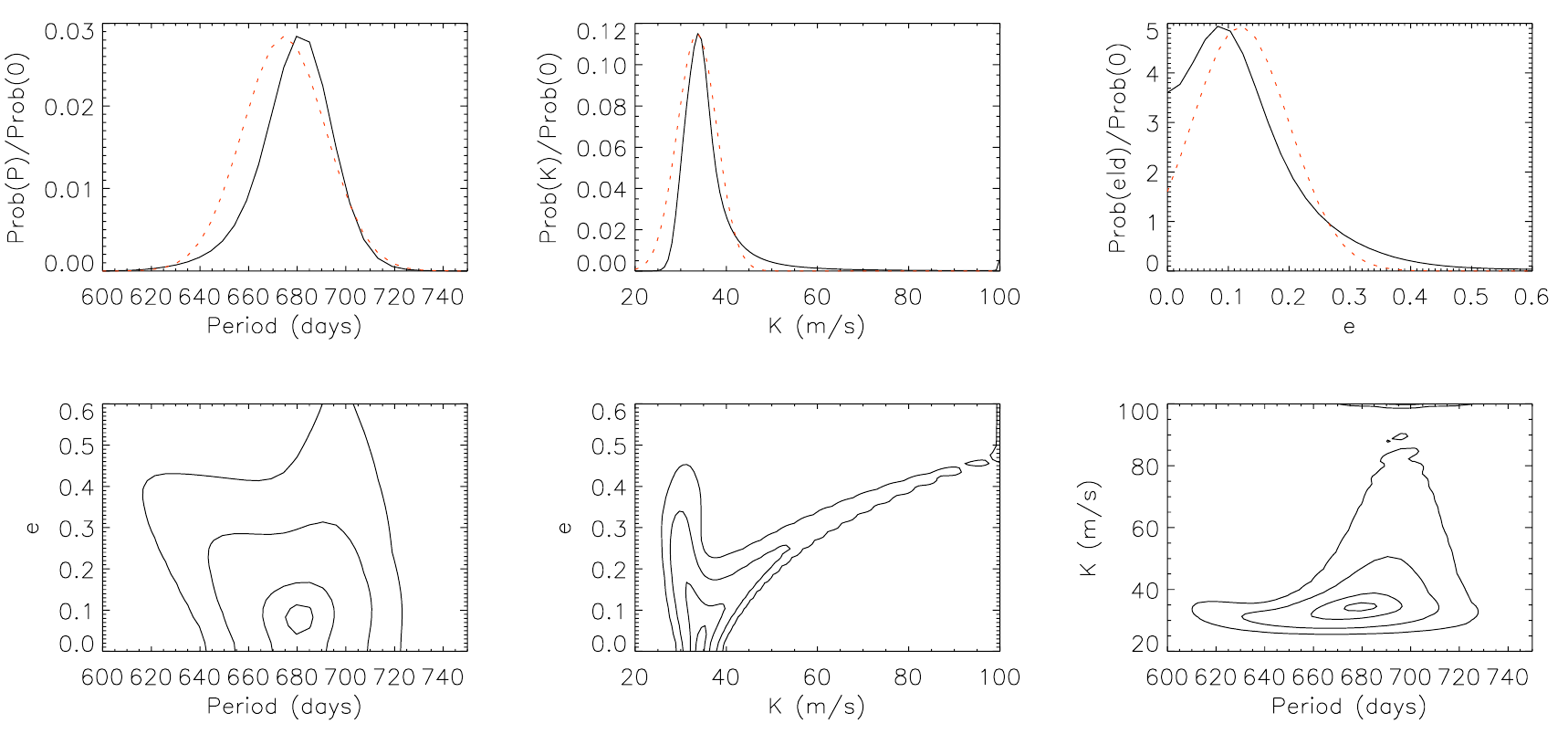}
\caption{Results of Keplerian fits to the 5319 data including a linear trend. 30 eccentricities, 30 periods and 100 $K$ values were calculated in the ranges shown. Dotted curves show Gaussian distributions with central values and standard deviations taken from \citealt{robinson}. Contours enclose 10\%, 50\%, 90\%, and 99\% of the total probability.\label{fig:5319ecc}}
\end{figure*}

\subsection{2D periodograms}

\cite{wright07} investigate the constraints that can be placed on the orbital parameters of long period orbits that have been only partially observed. They calculated the minimum $\chi^2$ at points across the $m\sin i$-$P$ plane. Similarly, \cite{otoole08} introduced a ``2D Keplerian Lomb-Scargle periodogram'' (2DKLS) in which the periodogram power is evaluated on a grid of $P$ and $e$, with a full Keplerian fit carried out at each point. \cite{otoole08} discuss the considerable computing resources being used to conduct simulations of detectability using this new 2D periodogram. The techniques we discuss earlier for rapid evaluation of multiple $\chi^2$ values could prove useful in more efficiently evaluating the 2DKLS periodogram. The constraints on $P$-$e$ or $P$-$K$ calculated in this paper differ from \cite{wright07} and \cite{otoole08} in that for each choice of $P$, $e$ or $P$, $K$ all values of the other parameters are taken into account, weighted by their probability, rather than finding the best fit values of the other parameters. This is the standard difference between Bayesian and frequentist approaches.

\cite{otoole08} mention that one of the reasons for looking at the periodogram power as a function of $P$ and $e$ is to help with detection of highly eccentric orbits. They consider the $e=0.97$ planet around HD20782 as an example. Their best fit has $e=0.97\pm 0.01$, $P=591.9\pm 2.8$, and $K=185.3\pm 49.7$. Our results for this data are shown in Figure \ref{fig:20782}. The discrete nature of the $K$ distribution is due to the finite sampling of the grid in eccentricity. The \cite{otoole08} solution lies on our contours, but towards the edge. The Bayesian calculation, which averages over the marginalized parameters, opens up a wider parameter space than the best-fit and error bars from \cite{otoole08} suggest.

\subsection{HD~5319}

HD 5319 has a planet with minimum mass $1.9\ M_J$ in a 675 day low eccentricity orbit \citep{robinson}. This is an interesting example to compare to because the analysis of \cite{robinson} used several different statistical methods. First, they used Monte Carlo simulations of data sets with noise only (simulated by selecting with replacement from the observed velocities) to assess the FAP, finding $1.3\times 10^{-3}$. They used both a scrambled velocity Monte Carlo simulations and an MCMC Bayesian calculation to estimate the uncertainties in the derived orbital parameters. They used an F-test to test the significance of including a linear trend in their model, finding a FAP of $3\times 10^{-4}$ indicating that a linear term is strongly preferred.

The results of our calculation are shown in Figure \ref{fig:5319ecc}. The dotted lines show the best fitting parameters and the errors found by \cite{robinson}, assuming Gaussian distributions, and agree well both in terms of central values and widths. Interestingly, the MCMC simulations run by \cite{robinson} did not agree as well with their scrambled velocity approach, whereas we find good agreement. The odds ratio for a trend in the no planet model is 0.9. For models with a planet, the odds ratios are $9.0\times 10^8$ (with trend) and $1.0\times 10^6$ (without trend). The model with a trend therefore has greater odds by a  factor of $10^3$, in good agreement with the F-test FAP of $3\times 10^{-4}$ found by \cite{robinson}. However, the overall false alarm probability we find $\sim 10^{-9}$ is much smaller than the simulations of \cite{robinson} suggested, $\sim 10^{-3}$.

\begin{figure}
\includegraphics[width=\columnwidth]{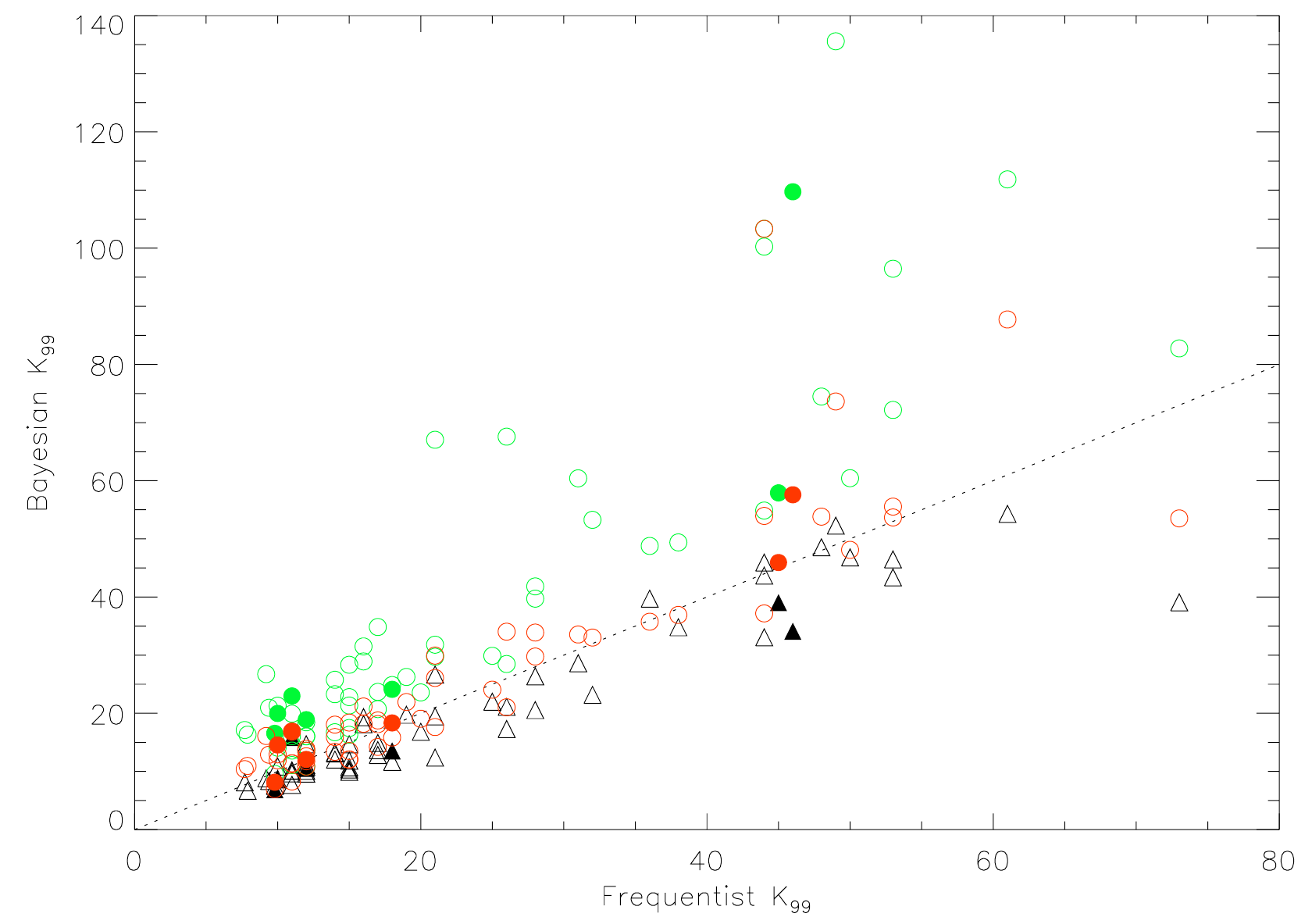}
\caption{Comparison between the 99\% upper limits on $K$ for circular orbits determined by \citealt{c99} for 63 stars from the Lick Planet Search, and the 99\% upper limits on $K$ from a Bayesian analysis of the same data. Solid triangles or circles include a linear trend in the fits (these are the 7 stars that \citealt{c99} found to have a significant slope), whereas open triangles or circles do not include a linear trend. The dotted line indicates a 1:1 correspondence between the two calculations of the upper limit. The black triangles are for circular orbits, the red circles are for eccentric orbits with $e<0.5$ and the green circles are for eccentric orbits with $e<0.7$.
\label{fig:Kup}}
\end{figure}

\subsection{Upper limits on $K$}

In an analysis of the Lick Planet Search, \cite{c99} calculated upper limits for 63 stars with non-detections. They used a Monte Carlo approach, in which simulated data sets with a circular orbit plus noise were analyzed and the velocity amplitude determined which resulted in detection 99\% of the time. We have reanalyzed the same data using our Bayesian scheme, first with circular orbits, and then with eccentric orbits. We calculate the 99\% upper limit $K_{99}$ by $\int_0^{K_{99}}\ dK\ \prob(K|d)=0.99$ (where $\prob(K|d)$ is normalized so that the total probability is unity).

The results are shown in Figure \ref{fig:Kup}. The triangles are for circular orbit models, and the circles are for eccentric orbits. For the 7 stars found to have a significant linear trend by \cite{c99}, we include a linear trend in the model. Overall the agreement is good. \cite{c99} (and \citealt{c08}) calculate upper limits for circular orbits to reduce the computational time needed. Based on the calculations of the effect of eccentricity on detectability of \cite{endl02} and \cite{c04}, they proposed that $K_{99}$ for circular orbits would be a good estimate of $K_{99}$ for orbits with $e\la 0.5$. We can test that here by calculating $K_{99}$ from the partial $K$ distribution
\begin{equation}
\prob(K|d)=\int^{e_{\rm cutoff}}_0\ de\ \prob(e,K|d)
\end{equation}
with different cutoffs $e_{\rm cutoff}$. For $e_{\rm cutoff}=0.9$, we find that the value of $K_{99}$ is generally much greater than $K_{99}$ for circular orbits, due to a tail of large $K$, large eccentricity solutions. However, for $e_{\rm cutoff}=0.5$, the agreement is very good. This is shown in Figure \ref{fig:Kup}, where we show results for $e_{\rm cutoff}=0.5$ (red symbols) and $e_{\rm cutoff}=0.7$ (green symbols). The $e_{\rm cutoff}=0.7$ values of $K_{99}$ are significantly greater than the circular orbit or $e_{\rm cutoff}=0.5$ values.


\section{Summary and Conclusions}

In this paper, we consider Bayesian analysis of radial velocity data. An advantage of this kind of analysis over traditional methods is that a single calculation gives the false alarm probability and the probability distributions of orbital period, eccentricity and velocity amplitude, allowing error bars or upper limits on these quantities to be determined. Using periodogram methods, separate calculations are required for each of these quantities, typically requiring many Monte Carlo trials.

Previous work on Bayesian analysis of radial velocities has used Markov Chain Monte Carlo (MCMC) techniques (although see \citealt{ford08} who used analytic techniques to partially carry out the marginalization for circular orbits). Our approach has been to apply some exact and approximate analytic results (based on previous work by \citealt{jaynes} and \citealt{bretthorst}) to the marginalization integrals for Keplerian fits to radial velocity data. In particular, we analytically integrate over the linear model parameters for each combination of $P$, $e$, and $t_p$, and use an analytic approximation (eq.~[\ref{eq:Kpdf}]) to reconstruct the probability distribution of $K$. An implementation of this algorithm in IDL is available on request from the authors.

With this approach, a full search of parameter space for a single Keplerian orbit takes several minutes on a $2$--$3\ {\rm GHz}$ processor, or several seconds for circular orbits, making it applicable to data sets from large velocity surveys. Constraints on orbital parameters (which involve surveying smaller regions of parameter space) can be calculated in seconds, competitive with MCMC techniques\footnote{In \cite{ford06}, the computer time needed was $\sim 10^{-6}\ {\rm s}\ N_{\rm obs}N_pL_cN_c$ where $N_{\rm obs}$ is the number of observations, $N_p$ the number of planets, $L_c$ the length of each chain, and $N_c$ the number of chains considered. For 30 observations, 1 planet, 10 chains each of length $10^4$ (multiple chains are required to assess convergence; \citealt{ford06}), the total time required is $\approx 3\ {\rm s}$.}. Our calculation can certainly be improved further. For example, we have focussed on the marginalization over the linear parameters in this paper, and used the simplest approach of evaluation on an evenly-spaced grid to integrate over the remaining parameters $P$, $K$, and $e$.

We compared our results with previous calculations. The constraints on orbital parameters and odds ratios agree well with MCMC results. We find that the Bayesian odds ratios are systematically lower than F-test odds ratios by a factor between 1 and 1000. This is due to the different accounting of trials in the two calculations \citep{c04}, with the Bayesian calculation including an Occam's razor penalty which accounts for the range of all parameters rather than only the frequency range. The techniques we have developed for rapidly calculating $\chi^2$ may have application to other techniques, such as the  2D periodograms of \cite{wright07} and \cite{otoole08}. We find good agreement with the upper limits on velocity amplitude $K$ calculated for circular orbits by \cite{c99} if we restrict our attention to $e\la 0.5$. More eccentric orbits give rise to a tail of solutions at large $K$. This shows that characterizing the $K$ distribution with a single parameter (e.g.~the 99\% upper limit; \citealt{c08}) is not appropriate for population analyses with highly eccentric orbits included. On the other hand, for low to moderate eccentricity orbits ($e\la 0.5$), upper limits can be derived from circular orbit fits which is much less numerically intensive.

The division of Keplerian parameters into ``fast'' and ``slow'' may prove useful in MCMC simulations. At the least, the systemic velocity does not need to be included as a parameter; it can be quickly evaluated for each set of the other parameters, and used to evaluate $\chi^2$ (this was also noted by \citealt{ford06}). One possible complication is that \cite{ford05} takes steps in a mixture of fast and slow parameters, $e\cos\omega$ and $e\sin\omega$, to help speed convergence. Separating the slow and fast parameters could potentially reduce efficiency in this case. Further investigations are needed.

\section*{Acknowledgments}

We thank Tyler Dodds for some early work on this problem during summer 2005, and Gil Holder for useful comments. AC acknowledges support from the National Sciences and Engineering Research Council of Canada (NSERC), Le Fonds Qu\'eb\'ecois de la Recherche sur la Nature et les Technologies (FQRNT), and the Canadian Institute for Advanced Research (CIFAR). AC is an Alfred P. Sloan Research Fellow.

\appendix
\section[]{Including a linear term (long term trend)}

In the main text, the ``no planet'' model that we have compared the sinusoid and Kepler fits to was a constant velocity model. Often, a linear term is included in the fit to account for long timescale trends in the data. Since adding a linear trend adds one extra linear term to the model, we can analytically marginalize over the slope in the same way as we marginalize over the constant term. In this Appendix, we give the formulae to do that.

\subsection{Is there evidence for a long term trend?}

First, consider a constant versus a linear model. Minimizing $\chi^2$ as a function of $\gamma$ for $V_i=\gamma$, we find the best fit constant term is
\begin{equation}
\gamma_{0}=\av{v},
\end{equation}
the corresponding minimum value of $\chi^2$ is 
\begin{equation}
{\chi^2_{\rm const}\over \sum w_i}=\avv{v^2},
\end{equation}
and
\begin{equation}
\det \alpha=\sum w_i.
\end{equation} 
Inserting these expressions into equation (\ref{eq:linearmarg}) with the number of parameters $m=1$ gives the posterior probability for a fit of a constant.

For a straight line fit, $V_i=\gamma+\beta t_i$, we find
\begin{eqnarray}
\gamma_0&=&{\av{v}\av{t^2}-\av{vt}\av{t}\over \avv{t^2}}\\
\beta_0&=&{\avv{vt}\over\avv{t^2}}\\
{\det\alpha\over (\sum w_i)^2}&=&\avv{t^2}\\
{\chi^2_{\rm line}\over \sum w_i}&=&\av{v^2}-2\gamma_0\av{v}-2\beta_0\av{vt}\nonumber\\&&+\gamma_0^2+2m\beta_0\gamma_0\av{t}+\beta_0^2\av{t^2}
\end{eqnarray}
Using equation (\ref{eq:linearmarg}), the odds ratio is
\begin{eqnarray}
\Lambda={(\chi^2_{\rm line})^{-(N-2)/2}\over (\chi^2_{\rm const})^{-(N-1)/2}}\left({\pi\over (\sum w_i)\avv{t^2}}\right)^{1/2}\nonumber\\{\Gamma((N-2)/2)\over \Gamma((N-1)/2)} {1\over \Delta \beta},
\end{eqnarray}
where $\Delta\beta$ is the prior range for $\beta$ (the prior range for $\gamma$ is the same in both models, and cancels). Here, we take $\beta$ to lie between $\pm \Delta v/T=\pm (v_{\rm max}-v_{\rm min})/T$, giving $\Delta\beta=2\Delta v/T$, that is we use the range of velocity amplitudes that we consider and the time of the observations to set the range of slopes. 

There is an important issue to mention here (we thank the referee for raising it), that the prior range of parameters should not depend on the data (the prior probability should reflect our state of knowledge before the data were taken). That is not true here since the range of observed velocities is used to determine what range of velocity amplitudes to search. Strictly, the normalization of the prior should not reflect this but be completely independent of the data. For example, the range of slopes could be set by looking at the range of slopes in previous planet discoveries (for example, in \citealt{butlercatalog} the reported slopes extend to $\approx 100\ {\rm m\ s^{-1}\ yr^{-1}}$), or the range of velocity amplitudes extend up to a maximum set by the amplitude induced by a $\approx 10\ M_J$ companion \cite{gregory05,fordgregory07}. However, the final odds ratios are not very sensitive to the exact choice of prior range. The range of velocity amplitudes enters the normalization logarithmically (since the prior is taken to be uniform in log). The range of slopes has the largest effect since it enters linearly, but we find that using a different choice, e.g. a range of $\beta$ from $-100$ to $+100 \ {\rm m\ s^{-1}\ yr^{-1}}$ changes the odds ratios by factors of a few to several only.

\subsection{Including a trend in the circular or Keplerian orbit fit}

Consider the model
\begin{equation}
V_i=\gamma+\beta t_i+A\sin\theta_i+B\cos\theta_i
\end{equation}
where $\theta_i=2\pi t_i/P$ for a circular orbit fit. Minimizing $\chi^2$ with respect to the four parameters $\gamma,\beta,A,B$, we find that their best fit values can be written in a concise way by defining a new average
\begin{equation}
\overline{xy}\equiv \avv{xy}-{\avv{xt}\avv{yt}\over\avv{t^2}}.
\end{equation}
Using this notation, 
\begin{eqnarray}\label{eq:m1}
\gamma_0&=&\av{v}-\beta_0\av{t}-A\av{S}-B\av{C}\\
\beta_0&=&{\avv{vt}-A\avv{St}-B\avv{Ct}\over \avv{t^2}}\\
A_0&=&{\avb{vS}\,\avb{C^2}-\avb{vC}\,\avb{SC}\over \avb{C^2}\,\avb{S^2}-\avb{SC}^2}\\
B_0&=&{\avb{vC}\,\avb{S^2}-\avb{vS}\,\avb{SC}\over \avb{C^2}\,\avb{S^2}-\avb{SC}^2}.
\end{eqnarray}
The expressions for $A_0$ and $B_0$ are the same as previously, but with the new averages. We also find
\begin{equation}
{\det\alpha\over \left(\sum w_i\right)^4}=\avv{t^2}\left[\avb{S^2}\,\avb{C^2}-\left(\avb{S}\,\avb{C}\right)^2\right]
\end{equation}
and
\begin{eqnarray}
{\chi^2_0\over \sum w_i}&=&\avv{v^2}-2\left(A_0\avv{vS}+B_0\avv{vC}\right)\nonumber\\&&+A_0^2\avv{S^2}+B_0^2\avv{C^2}+2A_0B_0\avv{SC}\nonumber\\ &&-\beta_0^2\avv{t^2}\nonumber\\ 
&=&\avb{v^2}-2\left(A_0\,\avb{vS}+B_0\,\avb{vC}\right)\nonumber\\&&+A_0^2\,\avb{S^2}+B_0^2\,\avb{C^2}+2A_0B_0\,\avb{SC}.\label{eq:m2}
\end{eqnarray}
Equations (\ref{eq:m1}) to (\ref{eq:m2}) replace equations (\ref{eq:c1}) to (\ref{eq:c2}) when a long term trend is included. They are essentially the same, but with the average $\avb{xy}$ used instead of $\avv{xy}$. Equation (\ref{eq:linearmarg}) with $m=4$ then allows marginalization over the four parameters $A,B,\beta$ and $\gamma$.

Similarly, for the grid based approach, the expression for $\chi^2$ is of the same form as equation (\ref{eq:chi2phikp}), but with the averages calculated as $\avb{xy}$ instead of $\avv{xy}$.

\section[]{Likelihood for fixed noise scaling parameter $k$}

In the main text, we integrated over the noise scaling parameter $k$, giving the likelihood in equation (\ref{eq:likelihood2}) (t-distribution) rather than equation (\ref{eq:likelihood}) (exponential). As we argued in \S 2.3, the analytic marginalization over an infinite range of $k$ is a good approximation for a reasonable spread in $k$. However, it could be that we are able to predict $k$ quite accurately, for example, if the level of stellar jitter has been predetermined for a particular star, in which case we might want to carry out a calculation for fixed $k$. Also, this would allow a calculation of the posterior probability for $k$. 

For fixed $k$, we have $\prob(d|\vec{a})\propto k^{-N} \exp(-\chi^2(\vec{a})/2k^2)$. The normalization over the constant term is then, taking circular orbits as an example, 
\begin{equation}
\prob(d|\phi,K,P)\propto \int^{\infty}_{-\infty}d\gamma\ k^{-N}\exp\left(-{\chi^2\over 2k^2}\right)
\end{equation}
where $\chi^2(\gamma)$ has the quadratic form of equation (\ref{eq:gammachi}). Therefore we can take
\begin{equation}
\prob(d|\phi,K,P)=k^{-(N-1)}\exp\left(-{\chi^2\left[\gamma_0,\phi,K,P\right]\over 2k^2}\right)
\end{equation}
as a replacement for equation (\ref{eq:probkpphi}), where we set the prefactor to unity as it cancels when we form the odds ratio. 

Similarly, the analytic result giving marginalization over $m$ parameters for a general linear model (eq.~[\ref{eq:linearmarg}]) becomes
\begin{eqnarray}
\int d^m\vec{a}\ k^{-N}\exp\left(-{\chi^2\over 2k^2}\right)\nonumber\\={(2\pi)^{m/2}k^{-(N-m)}\over \sqrt{\det\alpha}}\exp\left(-{\chi_0^2\over 2k^2}\right)
\end{eqnarray}
As a check, marginalization over $k$ at this stage takes us back to equation (\ref{eq:linearmarg}).

\label{lastpage}

\end{document}